\documentclass[aps,prapplied,reprint,superscriptaddress]{revtex4-1}

\usepackage{graphicx}
\usepackage{dcolumn}
\usepackage{bm}
\usepackage{hyperref}
\usepackage[mathlines]{lineno}
\usepackage{amsmath}
\usepackage{xfrac}
\usepackage{soul}
\usepackage{cancel}
\usepackage{xcolor}
\usepackage[nameinlink,capitalize]{cleveref}
\Crefname{figure}{Fig.}{Figs.}
\crefname{figure}{Fig.}{Figs.}

\crefname{equation}{Eq.}{Eqs.}
\Crefname{equation}{Eq.}{Eqs.}

\allowdisplaybreaks

\usepackage{xparse}
\ExplSyntaxOn
\NewDocumentCommand{\mref}{m}{\quinn_mref:n {#1}}
\seq_new:N \l_quinn_mref_seq
\cs_new:Npn \quinn_mref:n #1
 {
  \seq_set_split:Nnn \l_quinn_mref_seq { , } { #1 }
  \seq_pop_right:NN \l_quinn_mref_seq \l_tmpa_tl
  ( 
  \seq_map_inline:Nn \l_quinn_mref_seq
    { \ref{##1},\nobreakspace } 
  \exp_args:NV \ref \l_tmpa_tl 
  ) 
 }
\ExplSyntaxOff

\begin{document}

\preprint{APS/123-QED} 

\title{Capturing Complex Behaviour in Josephson Travelling Wave Parametric Amplifiers}

\author{T.~Dixon}
 
\affiliation{National Physical Laboratory, Hampton Road, Teddington, TW11 0LW, United Kingdom}
\affiliation{Royal Holloway University of London, Egham, Surrey, TW20 0EX, United Kingdom}%
\author{J.~W.~Dunstan}
\affiliation{Royal Holloway University of London, Egham, Surrey, TW20 0EX, United Kingdom}%
\author{G.~B.~Long}
\affiliation{National Physical Laboratory, Hampton Road, Teddington, TW11 0LW, United Kingdom}
\author{J.~M.~Williams}
\affiliation{National Physical Laboratory, Hampton Road, Teddington, TW11 0LW, United Kingdom}
\author{P.~J.~Meeson}
\affiliation{Royal Holloway University of London, Egham, Surrey, TW20 0EX, United Kingdom}%
\author{C.~D.~Shelly}
\email{connor.shelly@npl.co.uk}
\affiliation{National Physical Laboratory, Hampton Road, Teddington, TW11 0LW, United Kingdom}

\date{\today}

\begin{abstract}
We present an analysis of wave-mixing in recently developed Josephson Travelling Wave Parametric Amplifiers (JTWPAs). Circuit simulations performed using WRspice show the full behaviour of the JTWPA allowing propagation of all tones. The Coupled Mode Equations (CMEs) containing only pump, signal, and idler propagation are shown to be insufficient to completely capture complex mixing behaviour in the JTWPA. Extension of the CMEs through additional state vectors in the analytic solutions allows closer agreement with WRspice. We consider an ordered framework for the systematic inclusion of extended eigenmodes and make a qualitative comparison with WRspice at each step. The agreement between the two methods validates both approaches and provides insight into the operation of the JTWPA. We show that care should be taken when using the CMEs and propose that WRspice should be used as a design tool for non-linear superconducting circuits such as the JTWPA.
\end{abstract}

\maketitle


\section{\label{sec:introduction}introduction}

\subsection{\label{sec:JTWPA}Josephson Travelling Wave Parametric Amplifiers}

Josephson junction (JJ) based parametric amplifiers (JPAs) \cite{Yurke_PRA_1989,yamamoto_2008_APL_flux_JPA,mutus_white_aip_2014_couple_JPA} have been used in recent years to provide quantum-limited noise performance for quantum optics experiments \cite{Mallet_Beltran_quantum_optics_JPA}, single microwave photon detection \cite{lehnert_single_mw_photon_JPA}, high-fidelity qubit readout for quantum information technologies \cite{siddiqi_hacohen_cqed_jpa,lin_aip_2013_singleshot_qubit}, as well as producing squeezed states \cite{Castellanos-Beltran2008}. These microwave, small signal, amplifiers have been shown to exhibit large gain ($>20~$dB) \cite{beltran-reso,Zhou_Esteve_chalmers_APS_2014_High_gain_array}, and approach the quantum noise limit \cite{Teufel_NNano_2009}. Typically these amplifiers have utilised high-Q superconducting resonators which have a limited bandwidth and dynamic range. Removing the resonant architecture and allowing non-linear interactions along a transmission line can increase both the dynamic range and bandwidth \cite{Eom2012_dynamic_range}.  More recently, the Josephson Travelling Wave Parametric Amplifier (JTWPA), based on JJs embedded in a microwave transmission line has been shown to provide large gain without the bandwidth limitation of the JPAs \cite{Macklin_Science_2015,White_APL_2015,Yaakobi_PRB_2013}.

\begin{figure}[tb]
\includegraphics[width=\columnwidth]{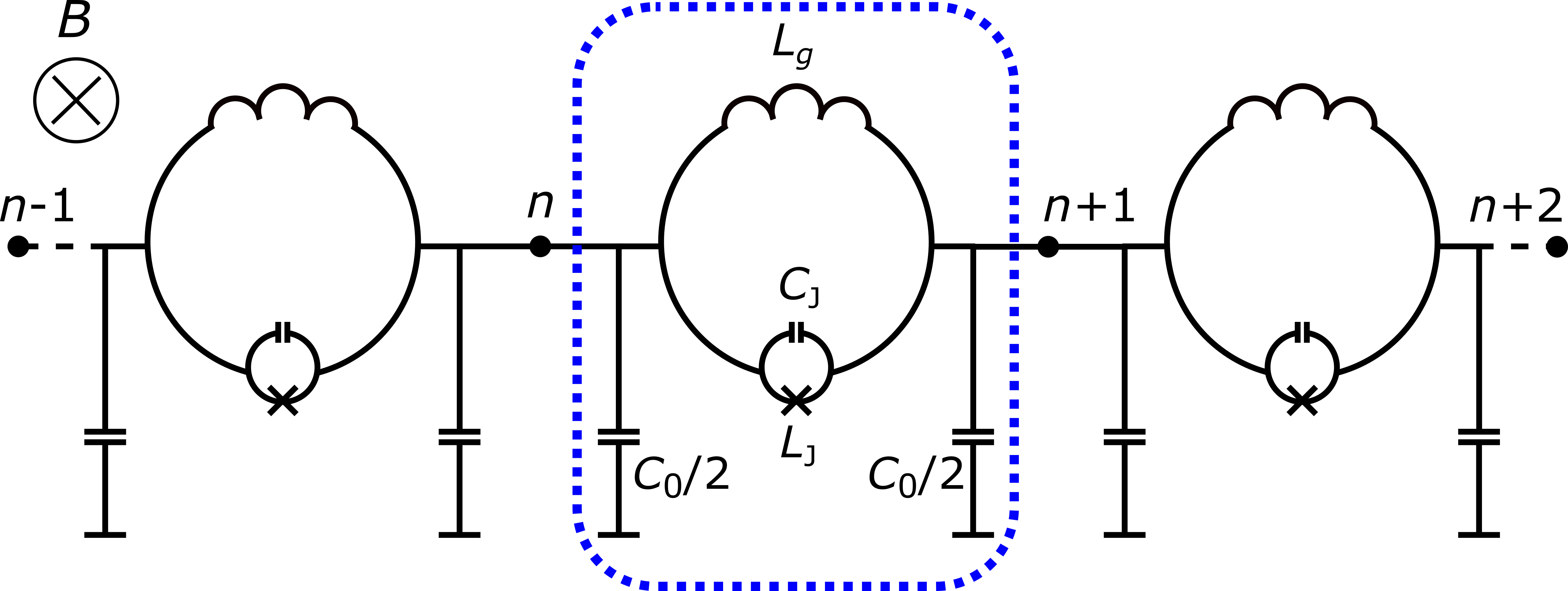}
\caption{Circuit schematic showing 3 cells of an N-cell array of rf-SQUIDs implemented in a microwave transmission line design as proposed by Zorin \cite{Zorin_PRAppl_2016}. The blue dotted line highlights one of the repeating cell elements. An externally applied magnetic field is used to shift the operation of the JTWPA to the 3WM regime (in the WRspice simulations, flux coupling to the rf-SQUID is used to bias the JTWPA). Each cell consists of an rf-SQUID with geometric inductance $L_{\mathrm{g}}$ and a Josephson junction (Josephson inductance $L_{\mathrm{J}}$, and junction capacitance $C_{\mathrm{J}}$). Each cell has a capacitance to ground $C_{\mathrm{0}}$. The parameters of the circuit presented in this work are $I_{\mathrm{c}}=5\,\mu\mathrm{A}$ $C_{\mathrm{J}}=60\,\mathrm{fF}$, $C_{\mathrm{0}}=100\,\mathrm{fF}$, and $L_{\mathrm{g}}=57\,\mathrm{pH}$.
\label{fig:cell}}
\end{figure}

Implementing unbiased JJs along the transmission line leads to a centrosymmetry of the system and four wave mixing (4WM) whereby the signal and idler frequencies are close to the frequency of the pump, $f_{\mathrm{s}}+f_{\mathrm{i}} = 2f_{\mathrm{p}}$. In this paper we focus primarily on the three wave mixing (3WM) scheme, $f_{\mathrm{s}}+f_{\mathrm{i}}=f_{\mathrm{p}}$ which shifts the pump frequency away from that of the signal and idler allowing the pump to be filtered more easily from the signal. The 3WM regime also takes advantage of the inherently stronger interactions than the 4WM regime. In this regime the phase modulation effect and the signal gain are controlled independently, the process of which is described in detail by Zorin \cite{Zorin_PRAppl_2016}. To access the 3WM regime rf-SQUIDs are embedded in the transmission line and an externally applied magnetic field modifies the centrosymmetry of the circuit. The circuit as proposed in  Ref.~\onlinecite{Zorin_PRAppl_2016} is shown in \Cref{fig:cell}.

\section{\label{sec:method}Modelling the JTWPA}

\subsection{\label{sec:WRspice}WRspice Simulations}

\begin{figure}[tb]
\includegraphics[width=\columnwidth]{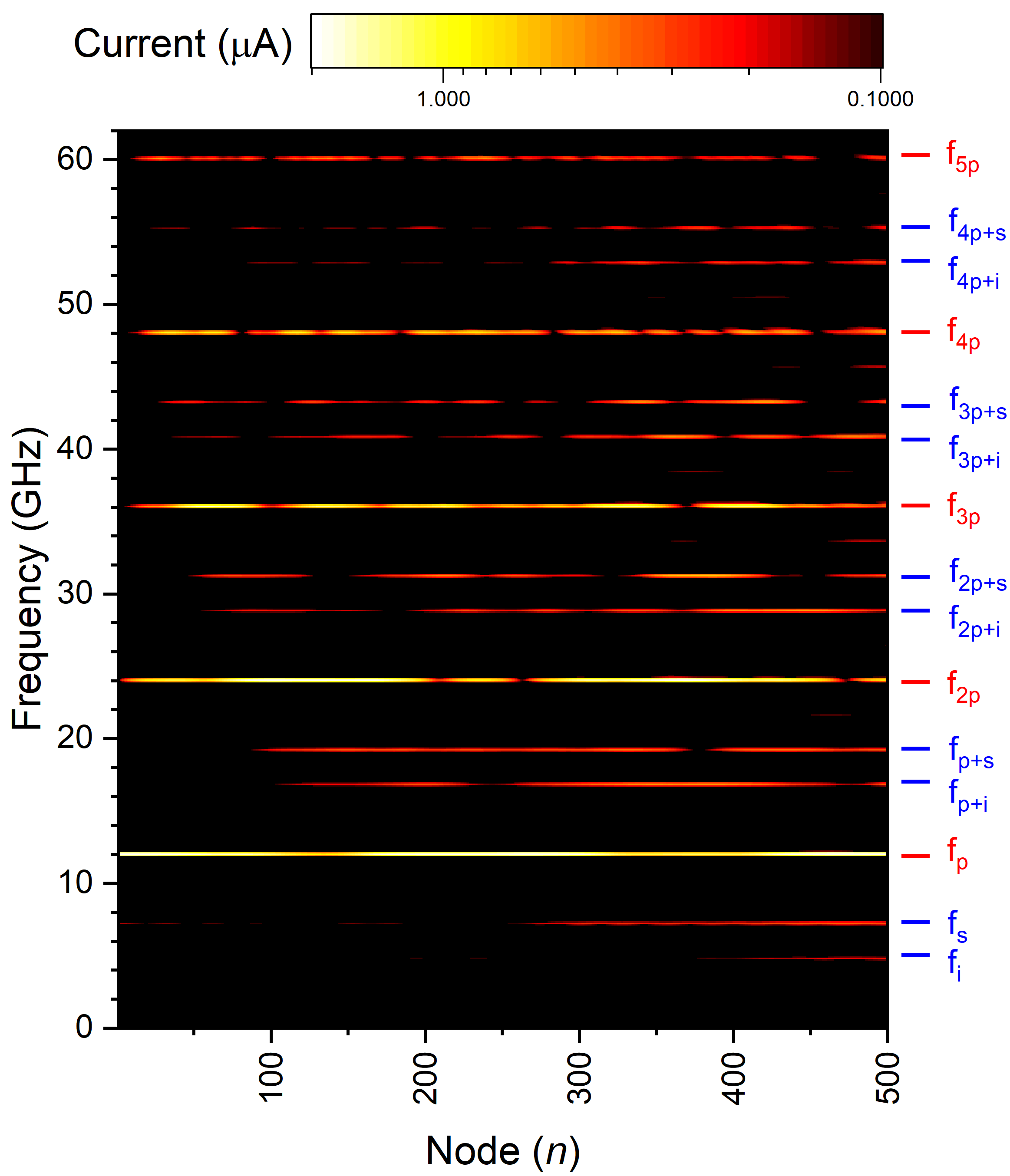}
\caption{Colourmap showing the current at each node of the JTWPA circuit, as simulated in WRspice. The colourmap shows the expected pump, signal, and idler tones as well as generation of additional tones. The harmonics of the pump (second-, third-, fourth- and fifth-harmonic generation shown) are clearly observed. Sum frequency generation (pump + idler, pump + signal, etc) are also observed. Pump harmonic terms labelled in red, pump-mediated sum-frequency generation labelled in blue. 
\label{fig:WRsim1}}
\end{figure}

In order to capture the full behaviour of the JTWPA, we use WRspice to simulate the circuit design shown in Fig.~\ref{fig:cell}. WRspice is a SPICE-like circuit simulator which includes a Josephson junction model \cite{WRspice}. Conventional analytical models describing three-wave mixers consider only three mixing tones, the pump $f_{\mathrm{p}}$, signal $f_{\mathrm{s}}$, and idler $f_{\mathrm{i}}$ \cite{Cullen_1960}. Using WRspice we observe that other mixing tones, especially the harmonics of the pump, are generated in the JTWPA. In this work we show that generation of other mixing tones has a strong reduction on the signal gain that can be achieved.  In WRspice we implement a 2000 cell version of the circuit shown in \Cref{fig:cell}. The rf-SQUIDs are flux-biased such that we operate in the 3WM regime. A strong ($I^{\mathrm{rms}}_{\mathrm{p}}(0)\approx1.97\,\mu\mathrm{A}\approx-70\,\mathrm{dBm}$), pump current at $f_{\mathrm{p}} = 12~$GHz and a weak ($I^{\mathrm{rms}}_{\mathrm{s}}(0)\approx0.07\,\mu\mathrm{A}\approx-96\,\mathrm{dBm}$) signal current at $f_{\mathrm{s}} = 7.2~$GHz are input to the JTWPA at node 0. The values replicate those used as example parameters in the analytical model by Zorin \cite{Zorin_PRAppl_2016}. By performing an FFT of the current entering each node $n$ we observe the behaviour of all tones propagating along the amplifier. We observe wave mixing processes including generation of the idler tone ($f_{\mathrm{i}} = 4.8~$GHz) at the difference of the pump and signal tones. This wave mixing derives solely from the non-linear current phase relation of the Josephson junction $I=I_{\mathrm{c}}\sin(\phi)$ and demonstrates the ability of WRspice to model the non-linear behaviour of the system. \Cref{fig:WRsim1} shows a colourmap of the current at each node of the JTWPA as simulated by WRspice. Note that as well as the signal, pump, and idler, we observe significant generation of pump harmonics $f_{\mathrm{2p}}$, $f_{\mathrm{3p}}$, $f_{\mathrm{4p}}$, and $f_{\mathrm{5p}}$. In addition to the pump harmonics, we observe sum-frequency generation associated with the pump and the pump harmonics.

The signal tone amplifies along the JTWPA from an input amplitude of $I^{\mathrm{rms}}_{\mathrm{s}}(0)\approx0.10\,\mu\mathrm{A}$ to {$I^{\mathrm{rms}}_{\mathrm{s}}(399)\approx0.19\,\mu\mathrm{A}$} representing a signal gain of 5.5 dB. This gain is less than a third of that predicted in Ref.~\onlinecite{Zorin_PRAppl_2016} for the same pump and signal input amplitudes and JTWPA length. We show here that the generation of the additional terms seen in the WRspice simulations accounts for most of the reduction in amplifier gain observed in WRspice when compared to the gain expected from the analytical theory described in Ref.~\onlinecite{Zorin_PRAppl_2016}. It is therefore clear that for the given circuit parameters additional tones must be taken into account in the analytical theory.

\subsection{\label{Numerical model}Extension of the Coupled Mode Equations}

To allow the analytical theory to capture more of the behaviour demonstrated by the JTWPA simulations we extend the coupled mode equations (CMEs) to include additional tones. The theory extension method is similar to that considered by Chaudhuri \textit{et al} for the 4WM case \cite{Chaudhuri_Gao_2015_4wm_CME_extend}. In \cref{tab:CME} the conventional theory as presented in Ref.~\onlinecite{Zorin_PRAppl_2016} is denoted as `CME-1' and includes the pump, signal and idler tones. Each further CME extension (CME-$k$) contains all pump-mediated mixing tones up to and including the $k^{\mathrm{th}}$-harmonic of the pump. Here we extend up to CME-5. The constituent tones of each CME set are shown in \cref{tab:CME}.
\begin{table}[h!]
\caption{\label{tab:CME}Tones included in each Coupled Mode Equation (CME) set. (CME-$k$) contains all pump-mediated mixing tones up to, and including the $k^{\mathrm{th}}$-harmonic of the pump.}
\begin{ruledtabular}
\begin{tabular}{lllll}
CME-1 & CME-2 & CME-3 & CME-4 & CME-5\\
\hline\\
$f_{\mathrm{i}}$ & $f_{\mathrm{i}}$ & $f_{\mathrm{i}}$ & $f_{\mathrm{i}}$ & $f_{\mathrm{i}}$\\
$f_{\mathrm{s}}$ & $f_{\mathrm{s}}$ & $f_{\mathrm{s}}$ & $f_{\mathrm{s}}$ & $f_{\mathrm{s}}$\\
$f_{\mathrm{p}}$ & $f_{\mathrm{p}}$ & $f_{\mathrm{p}}$ & $f_{\mathrm{p}}$ & $f_{\mathrm{p}}$\\
 & $f_{\mathrm{p+i}}$ & $f_{\mathrm{p+i}}$ & $f_{\mathrm{p+i}}$ & $f_{\mathrm{p+i}}$\\
& $f_{\mathrm{p+s}}$ & $f_{\mathrm{p+s}}$ & $f_{\mathrm{p+s}}$ & $f_{\mathrm{p+s}}$\\
& $f_{\mathrm{2p}}$ & $f_{\mathrm{2p}}$ & $f_{\mathrm{2p}}$  & $f_{\mathrm{2p}}$\\
&  & $f_{\mathrm{2p+i}}$ & $f_{\mathrm{2p+i}}$ & $f_{\mathrm{2p+i}}$\\
&  & $f_{\mathrm{2p+s}}$ & $f_{\mathrm{2p+s}}$ & $f_{\mathrm{2p+s}}$\\
&  & $f_{\mathrm{3p}}$ & $f_{\mathrm{3p}}$ & $f_{\mathrm{3p}}$\\
&  &  & $f_{\mathrm{3p+i}}$ & $f_{\mathrm{3p+i}}$\\
&  &  & $f_{\mathrm{3p+s}}$ & $f_{\mathrm{3p+s}}$\\
&  &  & $f_{\mathrm{4p}}$ & $f_{\mathrm{4p}}$\\
&  &  &  & $f_{\mathrm{4p+i}}$\\
&  &  &  & $f_{\mathrm{4p+s}}$\\
&  &  &  & $f_{\mathrm{5p}}$\\
\end{tabular}
\end{ruledtabular}
\end{table}

The inclusion of tones in the extended CMEs is described in detail below for the case of CME-2 (inclusion of the second harmonic of the pump, $f_\mathrm{2p}$, and the pump-mediated sum-frequency generations, $f_\mathrm{p+i}$ and $f_\mathrm{p+s}$). We introduce additional propagators $\partial A_{\mathrm{p+i}}/ \partial x$, $\partial A_{\mathrm{p+s}}/ \partial x$, and $\partial A_{\mathrm{2p}}/ \partial x$ in the allowed space of states $\Phi$ where, 
\begin{align}
\Phi = \sum_{j=\mathrm{i,s,p,p+i,p+s,2p}}A_{j}(x)e^{i(k_{j} x-\omega_{j} t)} + c.c.,\label{eq:SHGstates}
\end{align}
where $A_{j}(x)$ is the amplitude at dimensionless coordinate $x$ along the JTWPA of the $j^{\mathrm{th}}$-tone in the space of states. 

We treat these additional tones as a generated tone in the same way as the idler, that is, $A_{\mathrm{p+i}}(0)=A_{\mathrm{p+s}}(0)=A_{\mathrm{2p}}(0)=A_{\mathrm{i}}(0) = 0$. We then idealise our SQUID embedded transmission line to be purely non-centrosymmetric. This is the 3WM regime, where the coefficient of the cubic non-linearity $\gamma=0$. We follow the process outlined in Refs \onlinecite{Zorin_PRAppl_2016,Yaakobi_PRB_2013} to obtain the wave equation describing our transmission line of the form,
\begin{equation}
\begin{split}
    \dfrac{\partial ^2 \Phi}{\partial x^2} &- \omega_0^{-2} \dfrac{\partial^2 \Phi}{\partial t^2} + \omega_{\mathrm{J}}^{-2}\dfrac{\partial^4\Phi}{\partial x^2 \partial t^2} \\&+ \beta\dfrac{\partial}{\partial x}\Big[\Big( \dfrac{\partial \Phi}{\partial  x}\Big)^2 \Big]+ \cancelto{0}{\gamma\dfrac{\partial}{\partial x}\Big[\Big( \dfrac{\partial \Phi}{\partial  x}\Big)^3 \Big]} = 0, 
    \label{eq:TLwave}
\end{split}
\end{equation}
where,
\begin{align*}
    \omega_0 &= \frac{1}{\sqrt{L_{\mathrm{g}} C_0}}, 
    \quad\text{and}\quad 
    \omega_{\mathrm{J}} = \frac{1}{\sqrt{L_{\mathrm{g}} C_{\mathrm{J}}}},\\
\end{align*}
and,
\begin{align*}
\beta &= \beta_{\mathrm{L}}\frac{1}{2}\sin(\phi_{\mathrm{dc}}), \quad\text{and}\quad
    \beta_{\mathrm{L}} = \frac{2 \pi L_{\mathrm{g}} I_{\mathrm{c}}}{\Phi_0},\\
\end{align*}
where $\omega_{\mathrm{J}}$ is the plasma frequency, $\omega_{\mathrm{0}}$ is the cutoff frequency, with $L_{\mathrm{g}}$ the geometric inductance of the SQUID loop, $C_0$ capacitance to ground of the line, $C_{\mathrm{J}}$ the junction capacitance, $I_{\mathrm{c}}$ the junction critical current and $\Phi_0$ is the magnetic flux quantum. By assuming the non-linear component of \Cref{eq:TLwave} acts as a perturbation to the super-linear equation,
\begin{align}
\dfrac{\partial ^2 \Phi}{\partial x^2} - \omega_0^{-2} \dfrac{\partial^2 \Phi}{\partial t^2} + \omega_{\mathrm{J}}^2\dfrac{\partial^4\Phi}{\partial x^2 \partial t^2} = 0
\end{align}
we take the resulting super-linear dispersion solution,
\begin{align}
    k(\omega) = \frac{\omega}{\omega_0(\sqrt{1-\sfrac{\omega^2}{\omega_{\mathrm{J}}^2}})},
    \label{eq:kw}
\end{align}
and the space of allowed states in \cref{eq:SHGstates} as a trial solution to generate the coupled mode equations CME-2.

For frequencies much lower than the junction plasma frequency $\omega^{2}/\omega_{\mathrm{J}}^{2} \approx 0$, therefore \cref{eq:kw} can be simplified to $k(\omega)\approx \omega/\omega_{0}$. We now construct a simple set of CMEs including the tones $f_{\mathrm{p+i}}$, $f_{\mathrm{p+s}}$, and $f_{\mathrm{2p}}$ to find,
\begin{widetext}
\begin{align}
\dfrac{dA_{\mathrm{i}}}{dx} &= \dfrac{\beta}{2}\Big(k_{\mathrm{p}}k_{\mathrm{s}}A_{\mathrm{p}}A_{\mathrm{s}}^*e^{i(k_{\mathrm{p}}-k_{\mathrm{s}})x} + k_{\mathrm{p}}k_{\mathrm{p+i}}A_{\mathrm{p+i}}A_{\mathrm{p}}^*e^{i(k_{\mathrm{p+i}}-k_{\mathrm{p}})x}  + k_{\mathrm{2p}}k_{\mathrm{p+s}}A_{\mathrm{2p}}A_{\mathrm{p+s}}^*e^{i(k_{\mathrm{2p}}-k_{\mathrm{p+s}})x}  \Big)e^{-ik_{\mathrm{i}}x},\label{eq:CME2-i} \\
\dfrac{dA_{\mathrm{s}}}{dx} &= \dfrac{\beta}{2}\Big(k_{\mathrm{p}}k_{\mathrm{i}}A_{\mathrm{p}}A_{\mathrm{i}}^*e^{i(k_{\mathrm{p}}-k_{\mathrm{i}})x} + k_{\mathrm{p}}k_{\mathrm{p+s}}A_{\mathrm{p+s}}A_{\mathrm{p}}^*e^{i(k_{\mathrm{p+s}}-k_{\mathrm{p}})x}  + k_{\mathrm{2p}}k_{\mathrm{p+i}}A_{\mathrm{2p}}A_{\mathrm{p+i}}^*e^{i(k_{\mathrm{2p}}-k_{\mathrm{p+i}})x}  \Big)e^{-ik_{\mathrm{s}}x},\label{eq:CME2-s} \\
\dfrac{dA_{\mathrm{p}}}{dx} &=
\begin{aligned}[t]
\dfrac{\beta}{2}\Big(&-k_{\mathrm{s}}k_{\mathrm{i}}A_{\mathrm{i}}A_{\mathrm{s}}e^{i(k_{\mathrm{s}}+k_{\mathrm{i}})x} + k_{\mathrm{p+s}}k_{\mathrm{s}}A_{\mathrm{p+s}}A_{\mathrm{s}}^*e^{i(k_{\mathrm{p+s}}-k_{\mathrm{s}})x} + k_{\mathrm{p+i}}k_{\mathrm{i}}A_{\mathrm{p+i}}A_{\mathrm{i}}^*e^{i(k_{\mathrm{p+i}}-k_{\mathrm{i}})x}  \\  & 
+ k_{\mathrm{2p}}k_{\mathrm{p}}A_{\mathrm{2p}}A_{\mathrm{p}}^*e^{i(k_{\mathrm{2p}}-k_{\mathrm{p}})x} \Big)e^{-ik_{\mathrm{p}}x},\label{eq:CME2-p}
\end{aligned}\\
%
%
%
\dfrac{dA_{\mathrm{p+i}}}{dx} &= \dfrac{\beta}{2}\Big(-k_{\mathrm{p}}k_{\mathrm{i}}A_{\mathrm{p}}A_{\mathrm{i}}e^{i(k_{\mathrm{p}}+k_{\mathrm{i}})x} + k_{\mathrm{2p}}k_{\mathrm{s}}A_{\mathrm{2p}}A_{\mathrm{s}}^*e^{i(k_{\mathrm{2p}}-k_{\mathrm{s}})x} \Big)e^{-ik_{\mathrm{p+i}}x},\label{eq:CME2-pi} \\
\dfrac{dA_{\mathrm{p+s}}}{dx} &= \dfrac{\beta}{2}\Big(-k_{\mathrm{p}}k_{\mathrm{s}}A_{\mathrm{p}}A_{\mathrm{s}}e^{i(k_{\mathrm{p}}+k_{\mathrm{s}})x} + k_{\mathrm{2p}}k_{\mathrm{i}}A_{\mathrm{2p}}A_{\mathrm{i}}^*e^{i(k_{\mathrm{2p}}-k_{\mathrm{i}})x} \Big)e^{-ik_{\mathrm{p+s}}x},\label{eq:CME2-ps} \\
\dfrac{dA_{\mathrm{2p}}}{dx} &=\dfrac{\beta}{2}\Big(-\frac{k_{\mathrm{p}}^2A_{\mathrm{p}}^2}{2}e^{i(k_{\mathrm{p}}+k_{\mathrm{p}})x}
- k_{\mathrm{p+i}}k_{\mathrm{s}}A_{\mathrm{p+i}}A_{\mathrm{s}}e^{i(k_{\mathrm{p+i}}+k_{\mathrm{s}})x} - k_{\mathrm{p+s}}k_{\mathrm{i}}A_{\mathrm{p+s}}A_{\mathrm{i}}e^{i(k_{\mathrm{p+s}}+k_{\mathrm{i}})x} \Big)e^{-ik_{\mathrm{2p}}x}.\label{eq:CME2-pp}
\end{align}
\end{widetext}

Neglecting all terms proportional to $A_{\mathrm{p+i}}$, $A_{\mathrm{p+s}}$, and $A_{\mathrm{2p}}$, as well as  their derivatives shows that we recover the conventional CMEs used to describe the three wave parametric amplification,

\begin{align}
\dfrac{dA_{\mathrm{i}}}{dx} &= \dfrac{\beta}{2}\Big(k_{\mathrm{p}}k_{\mathrm{s}}A_{\mathrm{p}}A_{\mathrm{s}}^*e^{i(k_{\mathrm{p}}-k_{\mathrm{s}})x} \Big)e^{-ik_{\mathrm{i}}x} ,\label{eq:zi} \\
\dfrac{dA_{\mathrm{s}}}{dx} &= \dfrac{\beta}{2}\Big(k_{\mathrm{p}}k_{\mathrm{i}}A_{\mathrm{p}}A_{\mathrm{i}}^*e^{i(k_{\mathrm{p}}-k_{\mathrm{i}})x} \Big)e^{-ik_{\mathrm{s}}x},\label{eq:zS}\\
\dfrac{dA_{\mathrm{p}}}{dx} &=-\dfrac{\beta}{2}\Big(k_{\mathrm{s}}k_{\mathrm{i}}A_{\mathrm{s}}A_{\mathrm{i}}e^{i(k_{\mathrm{s}}+k_{\mathrm{i}})x} \Big)e^{-ik_{\mathrm{p}}x}\label{eq:zP}.
\end{align}
A similar set of extended equations are constructed for CMEs-3, 4, and 5 (see \crefrange{app:CME3}{app:CME5} for a full list of the equations). Each set of equations, CME-1 (\crefrange{eq:zi}{eq:zP}), CME-2 (\crefrange{eq:CME2-i}{eq:CME2-pp}), CME-3 (\crefrange{eq:CME3-p}{eq:CME3-2p+i}), CME-4 (\crefrange{eq:CME4-p}{eq:CME4-3p+i}), and CME-5 (\crefrange{eq:CME5-i}{eq:CME5-5p}) are solved numerically using the \texttt{ode45} function in MATLAB. 

\subsection{\label{sec:Comparison}Comparison of WRspice Simulations and Coupled Mode Equation Solutions}

In order to compare the WRspice simulations with the solutions to the Coupled Mode Equations it is necessary to relate the current $I(n)$ used in WRspice to the amplitude $A(x)$ used in the CMEs with the following relation,
\begin{equation}
I^{\mathrm{rms}}(n) =\lvert A(x)\rvert \frac{\omega L_{\mathrm{g}}I_{\mathrm{c}}}{\sqrt{2}\beta_{\mathrm{L}}Z},
\end{equation}
where $Z=\sqrt{L/C_{0}}$ is the impedance of the line.

\begin{figure}[tb]
\includegraphics[width=\columnwidth]{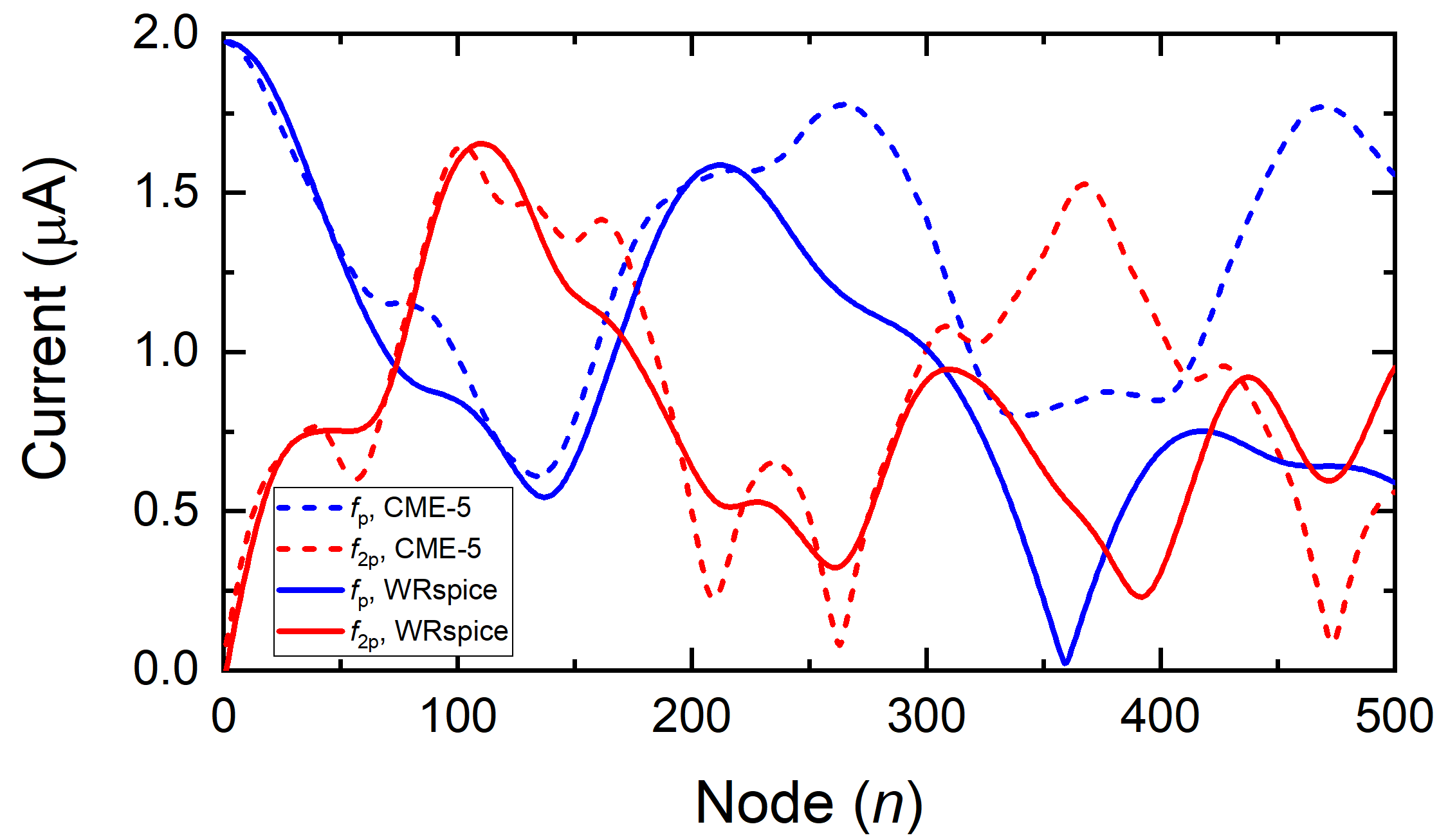}
\caption{Comparison between the extended CME-5 and the WRspice simulations of the pump ($f_{\mathrm{p}}$) and second harmonic of the pump ($f_{\mathrm{2p}}$). $I^{\mathrm{rms}}_{\mathrm{p}}(0)\approx1.97\,\mu\mathrm{A}$. The amplitude of both tones measured at each node as simulated in WRspice are well described using CME-5 up to node 250. 
\label{fig:SHG-fit}}
\end{figure}

To compare the WRspice simulations results with the solutions to the CMEs we focus first on the interaction between the pump $f_{\mathrm{p}}$ and the second harmonic of the pump $f_{\mathrm{2p}}$. From the WRspice output shown in \cref{fig:WRsim1} it is clear that the $f_{\mathrm{2p}}$ tone is of large amplitude and thus the second harmonic generation of the pump is a dominant mixing mechanism not accounted for in the CME-1 theory.  \cref{fig:SHG-fit} shows the solution to CME-5 for the  $f_{\mathrm{p}}$ and $f_{\mathrm{2p}}$ tones compared to the WRspice output. The amplitude of both tones is well described by the CME-5 solutions up to node 250, beyond which there is significant disagreement. 

\begin{figure}[tb]
\includegraphics[width=\columnwidth]{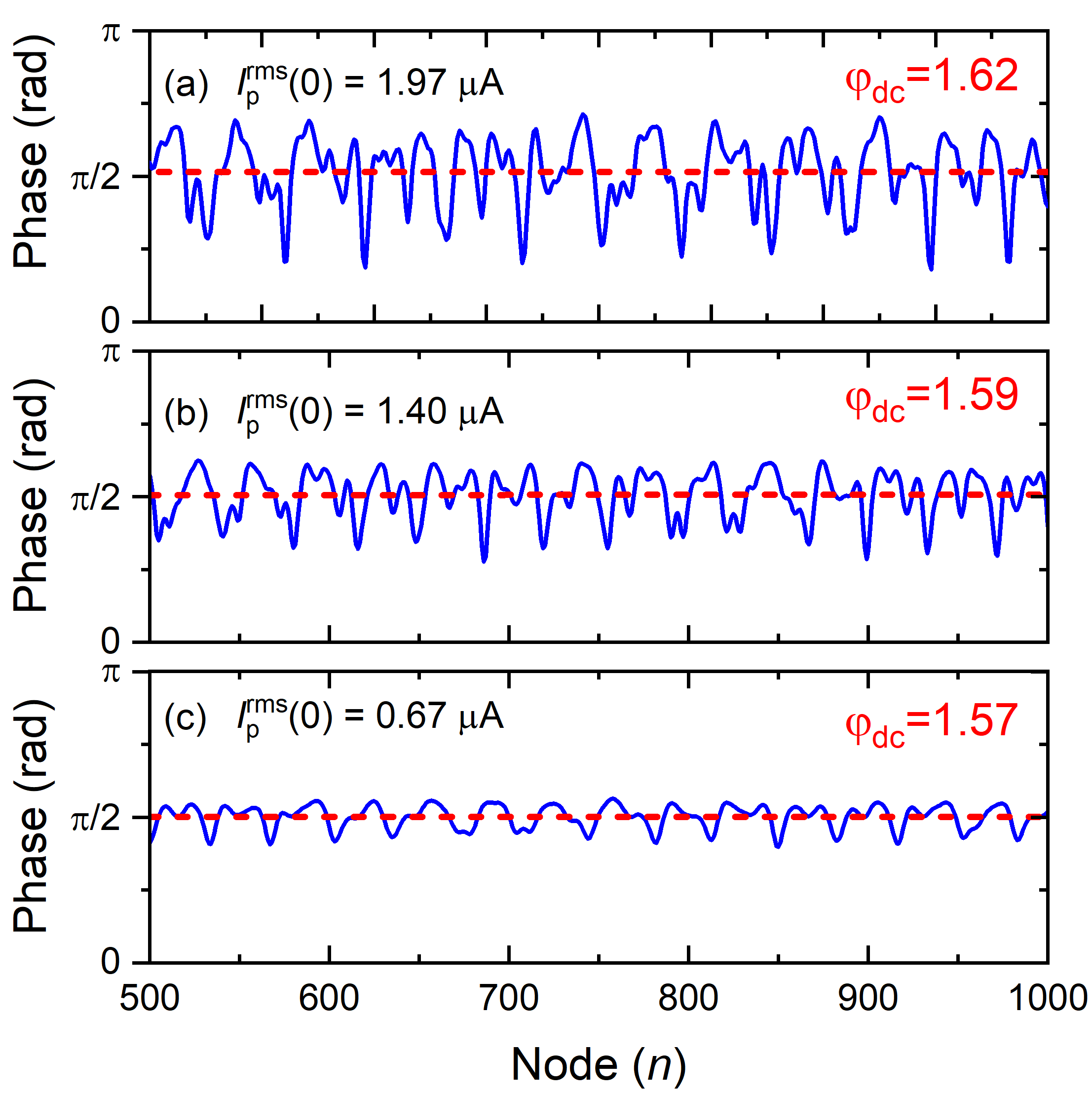}
\caption{Plots of WRspice junction phase between nodes 500 and 1000. Taken at $t=15\,\mathrm{ns}$. (a) $I^{\mathrm{rms}}_{\mathrm{p}}(0)\approx1.97\,\mu\mathrm{A}$ as per Ref \onlinecite{Zorin_PRAppl_2016}. Phase set to $\varphi_{\mathrm{dc}}=\pi/2$. Strong pump current causes large phase swing ($\approx\pm\pi/4$) and dc bias moves away from optimal position. (b) Pump current reduced to $I^{\mathrm{rms}}_{\mathrm{p}}(0)\approx1.40\,\mu\mathrm{A}$. Both phase swing and dc offset reduced. (c) Pump current used in simulations to investigate JTWPA signal gain $I^{\mathrm{rms}}_{\mathrm{p}}(0)\approx0.67\,\mu\mathrm{A}$. Minimal phase swing observed, dc bias position remaining at optimal position ($\varphi_{\mathrm{dc}}=\pi/2$).
\label{fig:phase}}
\end{figure}

There are a number of assumptions made in the original CME-1 theory (and carried through our CME extensions) that are now considered to ensure we are performing WRspice simulations in a regime in which these assumptions are broadly satisfied. The phase of the junction is set by a dc bias of $\varphi_{\mathrm{dc}}=\pi/2$ in order to operate in a purely non-centrosymmetric regime. The ac phase $\varphi_{\mathrm{ac}}$ is assumed to be small with respect to $\varphi_{\mathrm{dc}}$. \cref{fig:phase} shows that for high pump currents, approaching $2\,\mu\mathrm{A}$, $\varphi_{\mathrm{ac}}$ can no longer be considered to be small in comparison to $\varphi_{\mathrm{dc}}$. In addition, so called optical rectification is absent in the CMEs. Optical rectification is a dc offset generated by all other tones. The consequence of significant optical rectification is a deviation from the optimal $\varphi_{\mathrm{dc}}=\pi/2$ bias point such that the device no longer operates in the purely non-centrosymmetric regime. 

Note also that for such high input pump currents as shown in \cref{fig:WRsim1} for which $I^{\mathrm{rms}}_{\mathrm{p}}(0)\approx1.97\,\mu\mathrm{A}$, pump harmonic generation up to $f_{\mathrm{7p}}$ is observed (not shown in figure). As we only extend the CMEs to CME-5 we choose to reduce the input pump current such that pump harmonics beyond $f_{\mathrm{5p}}$ are insignificant, and that the assumption that $\varphi_{\mathrm{ac}}$ is small compared to $\varphi_{\mathrm{dc}}$ is upheld. \cref{fig:phase} shows that reducing the pump power from $I^{\mathrm{rms}}_{\mathrm{p}}(0)\approx1.97\,\mu\mathrm{A}$ to $I^{\mathrm{rms}}_{\mathrm{p}}(0)\approx0.67\,\mu\mathrm{A}$ reduces the amplitude of $\varphi_{\mathrm{ac}}$, and maintains a bias point $\varphi_{\mathrm{dc}}=1.57=\pi/2$ (non-centrosymmetric regime).

\begin{figure}[tb]
    \centering
    \includegraphics[width=\columnwidth]{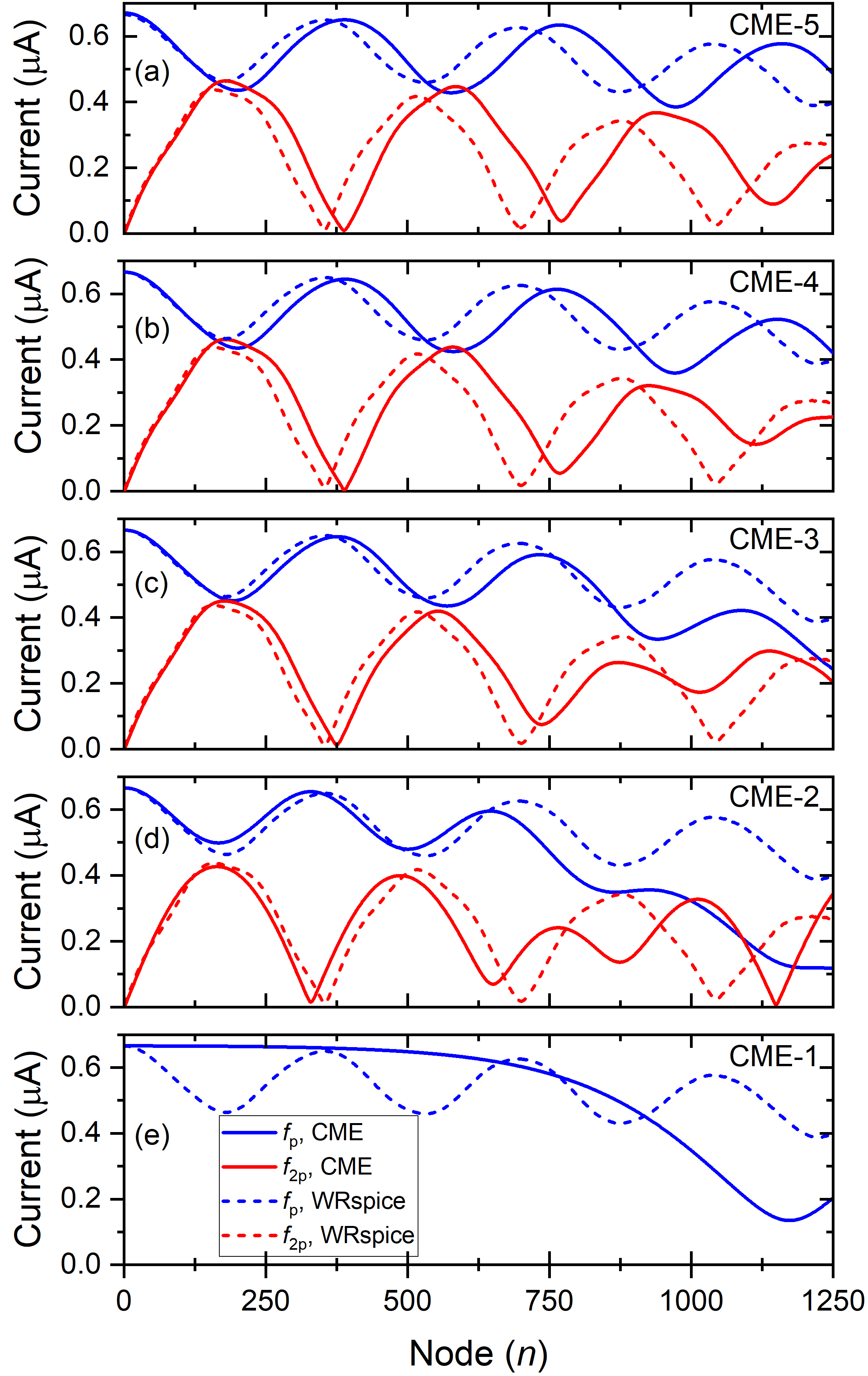}
    \caption{Pump, and the second harmonic of the pump currents as a function of node number. $I^{\mathrm{rms}}_{\mathrm{p}}(0)\approx0.67\,\mu\mathrm{A}$. The WRspice simulation of the pump and second harmonic of the pump are shown with dashed lines, and are the same for each panel. The {CME-$k$} solutions are shown with solid lines. Each panel shows a decreasing extension of CME. The agreement between the WRspice simulations and the CMEs reduces as the number of included tones in the CME are reduced. (a) CME-5, (b) CME-4, (c) CME-3, (d) CME-2, (e) CME-1.}
    \label{fig:SHG_inc}
\end{figure}

\cref{fig:SHG_inc} shows the current of the pump, and the current of the second harmonic of the pump along the JTWPA. The fit of CME-5 to the WRspice data is greatly improved with the pump power reduced to $I^{\mathrm{rms}}_{\mathrm{p}}(0)\approx0.67\,\mu\mathrm{A}$, and remains in agreement over more nodes. \cref{fig:SHG_inc} also shows that reducing the number of allowed states in the set of equations (i.e.,CME-5 $\rightarrow$ CME-4 $\rightarrow$ CME-3 $\rightarrow$ CME-2 $\rightarrow$ CME-1) results in an increased deviation of the agreement between the CME solutions and the WRspice output. These results show the risk of reducing the number of tones represented in the CME set. As the number of tones are reduced the behaviour of the pump and the second harmonic of the pump are less well described. Indeed, \cref{fig:SHG_inc}(d) shows no depletion of the pump due to second harmonic generation and that the CME-1 solutions do not capture the behaviour of the pump tone as simulated by WRspice.

\subsection{\label{Signal}Effect on Signal Gain}
\begin{figure}[tb]
\includegraphics[width=\columnwidth]{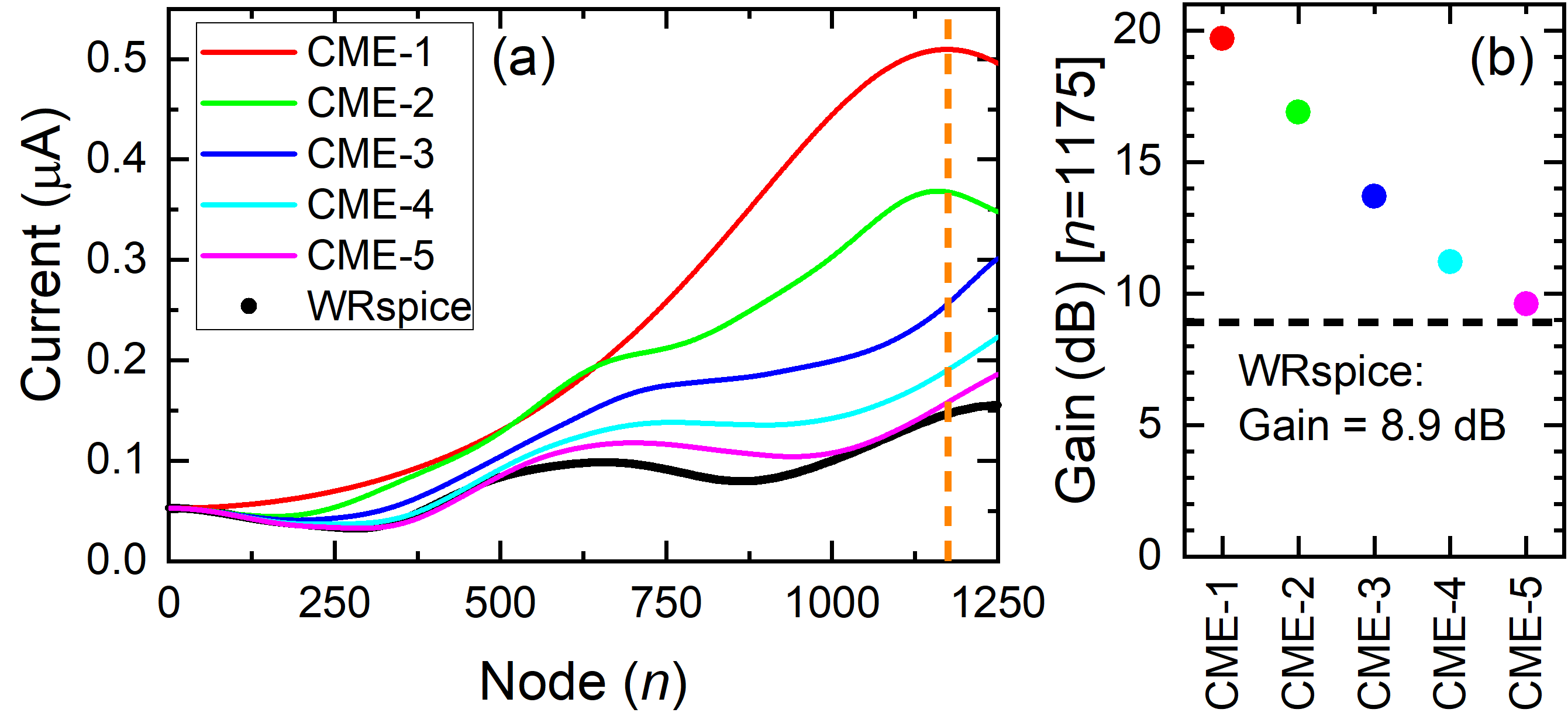}
\caption{(a) Signal current at each node of the JTWPA circuit for the WRspice simulations and each CME extension. With increasing CME extension we see improved agreement between the CME theory and the WRspice simulations. The traditional analytical theory CME-1 predicts a maximum signal at node $n=1175$ corresponding to a gain of 20 dB. We calculate the JTWPA gain for each CME set from the current measured at this node. (b) Gain measured at $n=1175$ for each CME extension. The WRspice simulation result is shown with a horizontal dashed line (G = 8.9 dB). The measured gain from each CME is reduced as the number of equations in the CME set increases. The gain measured approaches that of the value calculated from the WRspice simulations.   
\label{fig:signal_gain}}
\end{figure}

\cref{fig:signal_gain}(a) shows the signal current at each node of the JTWPA for the WRspice simulations and for CME-1 to CME-5. It can be seen that the presence of additional tones in the CMEs leads to a reduction in gain. CME-5 and WRspice are in fair agreement and exhibit the least gain.

To quantify the reduction in gain observed as the CMEs are extended, we choose the optimal gain node of CME-1 ($n=1175$) and compare to the other CMEs and the WRspice simulation at this node. \cref{fig:signal_gain}(b) shows as the number of terms in the CMEs increase we capture more complex behaviour of the signal as well as the detrimental effect on the gain. WRspice includes all tones propagating along the JTWPA, as noted earlier, and shows an even lower gain than CME-5 at node $n=1175$. 

\cref{fig:signal_gain}(a) also shows deamplification of the signal at the beginning of the JTWPA up to approximately node 300. We believe this deamplification is due to conservation of energy and the signal power dispersing into some of the other mixing tones. All tones, with the exception of the pump and the signal, are input to the equations with zero initial amplitude, and thus the power required to generate these tones must initially come from the pump and signal. It is observed that as the number of tones included in the CMEs increases, the number of nodes over which the signal deamplifies increases though the gradient is unchanged.

\section{Discussion and Conclusion}
Our extension of the CMEs show that CME-1 (including only the pump, signal, and idler) is insufficient to capture the complex behaviour of the JTWPA. As we increase the number of terms in the CMEs we approach the behaviour and gain figures observed in WRspice simulations. We note that whilst good agreement between CME-5 and WRspice is achieved, there is still not full agreement. We now speculate below on the sources of the remaining discrepancy. 

Only the quadratic term in the current-phase relation of the flux-biased SQUID is included in the formation of the CMEs. Inclusion of the quartic (and higher-order) terms may bring the WRspice and CME results into even better agreement. The dc offset generated by all other tones (optical rectification) is also not included in the CMEs whilst a dc current is seen in the WRspice for high pump currents. Finally, our choice of CME extensions are based on the WRspice results which show large amplitude pump harmonic and pump-mediated tones. Only these tones are included in the CME extensions we have presented in this work. Additional tones, including higher harmonics of the signal, may need consideration for improved agreement between WRspice and the CMEs.

We believe these results will have practical consequences for the design and operation of JTWPAs, in particular for considerations of measurement bandwidth, tone reflections,  and optimisation procedures.

To conclude, we demonstrate that a simple consideration of only three tones is insufficient to describe the complex behaviour of the JTWPA. We have presented four further extensions of the coupled mode equations, increasing the number of interacting tones included with each extension. We also used WRspice to simulate the JTWPA and compared its output to that of the extended coupled mode equations. Each further extension of the CMEs agreed more accurately with the WRspice simulation. 

We note that whilst good agreement between CME-5 and WRspice is achieved, there is still not full agreement and we have discussed possible reasons for this. In order to design an amplifier, and to obtain representative gain figures all of the behaviour of the JTWPA should be included. In this regard WRspice should be considered as the most reliable design tool. Both the simulations and the extended CME analytical theory show clearly that the generation of pump harmonics and the pump-mediated sum frequency generation terms must be considered when designing such a broadband device. In order to achieve the gains required for a usable JTWPA sufficient for quantum-limited amplification, engineering to suppress the pump harmonic generation may need be implemented. Some of this engineering is already considered in the form of stop-band engineering \cite{White_APL_2015,Zorin_PRAppl_2016,Zorin_flux_drive}.

This work realises a simple, computationally inexpensive, method for extension of the CMEs describing propagators which have been previously neglected and demonstrates the utility of WRspice for simulation of non-linear superconducting circuits, in particular as a design tool for JTWPAs.

\begin{acknowledgments}
  This project has received funding from the EMPIR programme co-financed by the Participating States and from the European Union’s Horizon 2020 research and innovation programme. This work is part of the the Joint Research Project PARAWAVE, and we would like to thank members of the consortium, in particular R.~Dolata, M.~Khabipov, C.~Ki{\ss}ling, and A.~B.~Zorin for useful discussions on the operation of the JTWPA. The work is partially supported by the UK Department of Business, Energy and Industrial Strategy (BEIS). We thank J.~Burnett and J.~C.~Gallop for critical review of the manuscript.
\end{acknowledgments}
\appendix
\begin{widetext}
\section{Extension to CME-3}
\label{app:CME3}
CME-3 extends CME-2 by including the third harmonic of the pump $f_{\mathrm{3p}}$, and the sum-frequency terms $f_{\mathrm{2p+i}}$ and $f_{\mathrm{2p+s}}$. We show in detail all of the terms included in the coupled mode equation forming CME-3.

\begin{align}
\dfrac{dA_{\mathrm{p}}}{dx} &=
\begin{aligned}[t]\dfrac{\beta}{2}\Big(
&-k_{\mathrm{s}}k_{\mathrm{i}}A_{\mathrm{s}}A_{\mathrm{i}}e^{i(k_{\mathrm{s}}+k_{\mathrm{i}})x}+k_{\mathrm{2p}}A_{\mathrm{2p}}k_{\mathrm{p}}A_{\mathrm{p}}^*e^{i(k_{\mathrm{2p}}-k_{\mathrm{p}})} 
+k_{\mathrm{p+s}}k_{\mathrm{s}}A_{\mathrm{p+s}}A_{\mathrm{s}}^*e^{i(k_{\mathrm{p+s}}-k_{\mathrm{s}})x} \\
& +k_{\mathrm{p+i}}k_{\mathrm{i}}A_{\mathrm{p+i}}A_{\mathrm{i}}^*e^{i(k_{\mathrm{p+i}}-k_{\mathrm{i}})x} + k_{\mathrm{3p}}k_{\mathrm{2p}}A_{\mathrm{3p}}A_{\mathrm{2p}}^*e^{i(k_{\mathrm{3p}}-k_{\mathrm{2p}})x} \\
& +k_{\mathrm{2p+s}}k_{\mathrm{p+s}}A_{\mathrm{2p+s}}A_{\mathrm{p+s}}^*e^{i(k_{\mathrm{2p+s}}-k_{\mathrm{p+s}})x} + k_{\mathrm{2p+i}}k_{\mathrm{p+i}}A_{\mathrm{2p+i}}A_{\mathrm{p+i}}^*e^{i(k_{\mathrm{2p+i}}-k_{\mathrm{p+i}})x}\Big)e^{-ik_{\mathrm{p}}x}
\label{eq:CME3-p}
\end{aligned}\\
\dfrac{dA_{\mathrm{s}}}{dx} &=
\begin{aligned}[t]\dfrac{\beta}{2}\Big(&k_{\mathrm{i}}k_{\mathrm{p}}A_{\mathrm{i}}^*A_{\mathrm{p}}e^{i(k_{\mathrm{p}}-k_{\mathrm{i}})x} + k_{\mathrm{p}}k_{\mathrm{p+s}}A_{\mathrm{p}}^*A_{\mathrm{p+s}}e^{i(k_{\mathrm{p+s}}-k_{\mathrm{p}})x} + k_{\mathrm{2p}}k_{\mathrm{p+i}}A_{\mathrm{p+i}}^*A_{\mathrm{2p}}e^{i(k_{\mathrm{2p}}-k_{\mathrm{p+i}})x} \\
&  +k_{\mathrm{3p}}k_{\mathrm{2p+i}}A_{\mathrm{3p}}A_{\mathrm{2p+i}}^*e^{i(k_{\mathrm{3p}}-k_{\mathrm{2p+i}})x}+k_{\mathrm{2p+s}}k_{\mathrm{2p}}A_{\mathrm{2p+s}}A_{\mathrm{2p}}^*e^{i(k_{\mathrm{2p+s}}-k_{\mathrm{2p}})x}\Big)e^{-ik_{\mathrm{s}}x}
\label{eq:CME3-s}
\end{aligned}\\
\dfrac{dA_{\mathrm{i}}}{dx} &=
\begin{aligned}[t]\dfrac{\beta}{2}\Big(&k_{\mathrm{s}}k_{\mathrm{p}}A_{\mathrm{s}}^*A_{\mathrm{p}}e^{i(k_{\mathrm{p}}-k_{\mathrm{s}})x}+k_{\mathrm{p}}k_{\mathrm{p+i}}A_{\mathrm{p}}^*A_{\mathrm{p+i}}e^{i(k_{\mathrm{p+i}}-k_{\mathrm{p}})x}+k_{\mathrm{2p}}k_{\mathrm{p+s}}A_{\mathrm{p+s}}^*A_{\mathrm{2p}}e^{i(k_{\mathrm{2p}}-k_{\mathrm{p+s}})x}  \\
&  +k_{\mathrm{3p}}k_{\mathrm{2p+s}}A_{\mathrm{3p}}A_{\mathrm{2p+s}}^*e^{i(k_{\mathrm{3p}}-k_{\mathrm{2p+s}})x}+k_{\mathrm{2p+i}}k_{\mathrm{2p}}A_{\mathrm{2p+i}}A_{\mathrm{2p}}^*e^{i(k_{\mathrm{2p+i}}-k_{\mathrm{2p}})x}\Big)e^{-ik_{\mathrm{i}}x}
\label{eq:CME3-i}
\end{aligned}\\
\dfrac{dA_{\mathrm{2p}}}{dx} &=
\begin{aligned}[t]\dfrac{\beta}{4}\Big(&k_{\mathrm{p}}^2A_{\mathrm{p}}^2e^{i(2k_{\mathrm{p}})x}-k_{\mathrm{i}}k_{\mathrm{p+s}}A_{\mathrm{i}}A_{\mathrm{p+s}}e^{i(k_{\mathrm{i}}+k_{\mathrm{p+s}})x}-k_{\mathrm{p+i}}k_{\mathrm{s}}A_{\mathrm{p+i}}A_{\mathrm{s}}e^{i(k_{\mathrm{p+i}}+k_{\mathrm{s}})x}\\
&  +k_{\mathrm{p}}k_{\mathrm{3p}}A_{\mathrm{p}}^*A_{\mathrm{3p}}e^{i(k_{\mathrm{3p}}-k_{\mathrm{p}})x}+k_{\mathrm{2p+s}}k_{\mathrm{s}}A_{\mathrm{2p+s}}A_{\mathrm{s}}^*e^{i(k_{\mathrm{2p+s}}-k_{\mathrm{s}})x} \\
&+ k_{\mathrm{2p+i}}k_{\mathrm{i}}A_{\mathrm{2p+i}}A_{\mathrm{i}}^*e^{i(k_{\mathrm{2p+i}}-k_{\mathrm{i}})x}\Big)e^{-ik_{\mathrm{2p}}x}
\label{eq:CME3-2p}
\end{aligned}\\
\dfrac{dA_{\mathrm{p+s}}}{dx} &=
\begin{aligned}[t]\dfrac{\beta}{2}\Big(&-k_{\mathrm{p}}k_{\mathrm{s}}A_{\mathrm{p}}A_{\mathrm{s}}e^{i(k_{\mathrm{p}}+k_{\mathrm{s}})x}+k_{\mathrm{i}}k_{\mathrm{2p}}A_{\mathrm{i}}^*A_{\mathrm{2p}}e^{i(k_{\mathrm{2p}}-k_{\mathrm{i}})x}    \\
&  +k_{\mathrm{3p}}k_{\mathrm{p+i}}A_{\mathrm{3p}}A_{\mathrm{p+i}}^*e^{i(k_{\mathrm{3p}}-k_{\mathrm{p+i}})x}+k_{\mathrm{2p+s}}k_{\mathrm{p}}A_{\mathrm{2p+s}}A_{\mathrm{p}}^*e^{i(k_{\mathrm{2p+s}}-k_{\mathrm{p}})x}\Big)e^{-ik_{\mathrm{p+s}}x}
\label{eq:CME3-p+s}
\end{aligned}\\
\dfrac{dA_{\mathrm{p+i}}}{dx} &=
\begin{aligned}[t]\dfrac{\beta}{2}\Big(&-k_{\mathrm{p}}k_{\mathrm{i}}A_{\mathrm{p}}A_{\mathrm{i}}e^{i(k_{\mathrm{p}}+k_{\mathrm{i}})x}+k_{\mathrm{s}}k_{\mathrm{2p}}A_{\mathrm{s}}^*A_{\mathrm{2p}}e^{i(k_{\mathrm{2p}}-k_{\mathrm{s}})x}    \\
&  +k_{\mathrm{3p}}k_{\mathrm{p+s}}A_{\mathrm{3p}}A_{\mathrm{p+s}}^*e^{i(k_{\mathrm{3p}}-k_{\mathrm{p+s}})x}+k_{\mathrm{2p+i}}k_{\mathrm{p}}A_{\mathrm{2p+i}}A_{\mathrm{p}}^*e^{i(k_{\mathrm{2p+i}}-k_{\mathrm{p}})x}\Big)e^{-ik_{\mathrm{p+i}}x}
\label{eq:CME3-p+i}
\end{aligned}\\
\dfrac{dA_{\mathrm{3p}}}{dx} &=
\begin{aligned}[t]\dfrac{\beta}{2}\Big(&-k_{\mathrm{p}}k_{\mathrm{2p}}A_{\mathrm{p}}A_{\mathrm{2p}}e^{i(k_{\mathrm{p}}+k_{\mathrm{2p}})x}-k_{\mathrm{p+s}}k_{\mathrm{p+i}}A_{\mathrm{p+s}}A_{\mathrm{p+i}}e^{i(k_{\mathrm{p+s}}+k_{\mathrm{p+i}})x}    \\
&  -k_{\mathrm{s}}k_{\mathrm{2p+i}}A_{\mathrm{s}}A_{\mathrm{2p+i}}e^{i(k_{\mathrm{s}}+k_{\mathrm{2p+i}})x}-k_{\mathrm{i}}k_{\mathrm{2p+s}}A_{\mathrm{i}}A_{\mathrm{2p+s}}e^{i(k_{\mathrm{i}}+k_{\mathrm{2p+s}})x}\Big)e^{-ik_{\mathrm{3p}}x}
\label{eq:CME3-3p}
\end{aligned}\\
\dfrac{dA_{\mathrm{2p+s}}}{dx} &=
\begin{aligned}[t]\dfrac{\beta}{2}\Big(&-k_{\mathrm{p}}k_{\mathrm{p+s}}A_{\mathrm{p}}A_{\mathrm{p+s}}e^{i(k_{\mathrm{p}}+k_{\mathrm{p+s}})x}-k_{\mathrm{s}}k_{\mathrm{2p}}A_{\mathrm{s}}A_{\mathrm{2p}}e^{i(k_{\mathrm{s}}+k_{\mathrm{2p}})x}\\
&+k_{\mathrm{3p}}k_{\mathrm{i}}A_{\mathrm{3p}}A_{\mathrm{i}}^*e^{i(k_{\mathrm{3p}}-k_{\mathrm{i}})x}\Big)e^{-ik_{\mathrm{2p+s}}x}
\label{eq:CME3-2p+s}
\end{aligned}\\
\dfrac{dA_{\mathrm{2p+i}}}{dx} &=
\begin{aligned}[t]\dfrac{\beta}{2}\Big(&-k_{\mathrm{p}}k_{\mathrm{p+i}}A_{\mathrm{p}}A_{\mathrm{p+i}}e^{i(k_{\mathrm{p}}+k_{\mathrm{p+i}})x}-k_{\mathrm{i}}k_{\mathrm{2p}}A_{\mathrm{i}}A_{\mathrm{2p}}e^{i(k_{\mathrm{i}}+k_{\mathrm{2p}})x}\\
&+k_{\mathrm{3p}}k_{\mathrm{s}}A_{\mathrm{3p}}A_{\mathrm{s}}^*e^{i(k_{\mathrm{3p}}-k_{\mathrm{s}})x}\Big)e^{-ik_{\mathrm{2p+i}}x}
\label{eq:CME3-2p+i}
\end{aligned}
\end{align}
\end{widetext}

\begin{widetext}
\section{Extension to CME-4}
\label{app:CME4}
The penultimate coupled mode equation extension that we present in full is the extension from CME-3 to CME-4 by inclusion of the fourth harmonic of the pump $f_{\mathrm{4p}}$, and the sum-frequency terms $f_{\mathrm{3p+i}}$ and $f_{\mathrm{3p+s}}$. The full list of tones included in CME-4 is shown in \cref{tab:CME}. We show below in detail all of the terms included in the coupled mode equation forming CME-4.

\begin{align}
\dfrac{dA_{\mathrm{p}}}{dx} &=
\begin{aligned}[t]\dfrac{\beta}{2}\Big(
&-k_{\mathrm{s}}k_{\mathrm{i}}A_{\mathrm{s}}A_{\mathrm{i}}e^{i(k_{\mathrm{s}}+k_{\mathrm{i}})x}+k_{\mathrm{2p}}A_{\mathrm{2p}}k_{\mathrm{p}}A_{\mathrm{p}}^*e^{i(k_{\mathrm{2p}}-k_{\mathrm{p}})x}+k_{\mathrm{p+s}}k_{\mathrm{s}}A_{\mathrm{p+s}}A_{\mathrm{s}}^*e^{i(k_{\mathrm{p+s}}-k_{\mathrm{s}})x} \\
& +k_{\mathrm{p+i}}k_{\mathrm{i}}A_{\mathrm{p+i}}A_{\mathrm{i}}^*e^{i(k_{\mathrm{p+i}}-k_{\mathrm{i}})x}+k_{\mathrm{3p}}k_{\mathrm{2p}}A_{\mathrm{3p}}A_{\mathrm{2p}}^*e^{i(k_{\mathrm{3p}}-k_{\mathrm{2p}})x}  \\
& +k_{\mathrm{2p+s}}k_{\mathrm{p+s}}A_{\mathrm{2p+s}}A_{\mathrm{p+s}}^*e^{i(k_{\mathrm{2p+s}}-k_{\mathrm{p+s}})x}+ k_{\mathrm{2p+i}}k_{\mathrm{p+i}}A_{\mathrm{2p+i}}A_{\mathrm{p+i}}^*e^{i(k_{\mathrm{2p+i}}-k_{\mathrm{p+i}})x}\\ 
&+ k_{\mathrm{4p}}k_{\mathrm{3p}}A_{\mathrm{4p}}A_{\mathrm{3p}}^*e^{i(k_{\mathrm{4p}}-k_{\mathrm{3p}})x} + k_{\mathrm{3p+s}}k_{\mathrm{2p+s}}A_{\mathrm{3p+s}}A_{\mathrm{2p+s}}^*e^{i(k_{\mathrm{3p+s}}-k_{\mathrm{2p+s}})x}\\
&+ k_{\mathrm{3p+i}}k_{\mathrm{2p+i}}A_{\mathrm{3p+i}}A_{\mathrm{2p+i}}^*e^{i(k_{\mathrm{3p+i}}-k_{\mathrm{2p+i}})x}\Big)e^{-ik_{\mathrm{p}}x}
\label{eq:CME4-p}
\end{aligned}\\
\dfrac{dA_{\mathrm{s}}}{dx} &=
\begin{aligned}[t]\dfrac{\beta}{2}\Big(&k_{\mathrm{i}}k_{\mathrm{p}}A_{\mathrm{i}}^*A_{\mathrm{p}}e^{i(k_{\mathrm{p}}-k_{\mathrm{i}})x}+k_{\mathrm{p}}k_{\mathrm{p+s}}A_{\mathrm{p}}^*A_{\mathrm{p+s}}e^{i(k_{\mathrm{p+s}}-k_{\mathrm{p}})x}+k_{\mathrm{2p}}k_{\mathrm{p+i}}A_{\mathrm{p+i}}^*A_{\mathrm{2p}}e^{i(k_{\mathrm{2p}}-k_{\mathrm{p+i}})x} \\
&+k_{\mathrm{3p}}k_{\mathrm{2p+i}}A_{\mathrm{3p}}A_{\mathrm{2p+i}}^*e^{i(k_{\mathrm{3p}}-k_{\mathrm{2p+i}})x}+k_{\mathrm{2p+s}}k_{\mathrm{2p}}A_{\mathrm{2p+s}}A_{\mathrm{2p}}^*e^{i(k_{\mathrm{2p+s}}-k_{\mathrm{2p}})x}\\
&+k_{\mathrm{4p}}k_{\mathrm{3p+i}}A_{\mathrm{4p}}A_{\mathrm{3p+i}}^*e^{i(k_{\mathrm{4p}}-k_{\mathrm{3p+i}})x} + k_{\mathrm{3p}}k_{\mathrm{3p+s}}A_{\mathrm{3p+s}}A_{\mathrm{3p}}^*e^{i(k_{\mathrm{3p+s}}-k_{\mathrm{3p}})x}\Big)e^{-ik_{\mathrm{s}}x}
\label{eq:CME4-s}
\end{aligned}\\
\dfrac{dA_{\mathrm{i}}}{dx} &=
\begin{aligned}[t]\dfrac{\beta}{2}\Big(&k_{\mathrm{s}}k_{\mathrm{p}}A_{\mathrm{s}}^*A_{\mathrm{p}}e^{i(k_{\mathrm{p}}-k_{\mathrm{s}})x}+k_{\mathrm{p}}k_{\mathrm{p+i}}A_{\mathrm{p}}^*A_{\mathrm{p+i}}e^{i(k_{\mathrm{p+i}}-k_{\mathrm{p}})x}+k_{\mathrm{2p}}k_{\mathrm{p+s}}A_{\mathrm{p+s}}^*A_{\mathrm{2p}}e^{i(k_{\mathrm{2p}}-k_{\mathrm{p+s}})x} \\
&  +k_{\mathrm{3p}}k_{\mathrm{2p+s}}A_{\mathrm{3p}}A_{\mathrm{2p+s}}^*e^{i(k_{\mathrm{3p}}-k_{\mathrm{2p+s}})x}+k_{\mathrm{2p+i}}k_{\mathrm{2p}}A_{\mathrm{2p+i}}A_{\mathrm{2p}}^*e^{i(k_{\mathrm{2p+i}}-k_{\mathrm{2p}})x}\\
&+ k_{\mathrm{4p}}k_{\mathrm{3p+s}}A_{\mathrm{4p}}A_{\mathrm{3p+s}}^*e^{i(k_{\mathrm{4p}}-k_{\mathrm{3p+s}})x} + k_{\mathrm{3p}}k_{\mathrm{3p+i}}A_{\mathrm{3p+i}}A_{\mathrm{3p}}^*e^{i(k_{\mathrm{3p+i}}-k_{\mathrm{3p}})x}\Big)e^{-ik_{\mathrm{i}}x}
\label{eq:CME4-i}
\end{aligned}\\
\dfrac{dA_{\mathrm{2p}}}{dx} &=
\begin{aligned}[t]\dfrac{\beta}{2}\Big(&\dfrac{k_{\mathrm{p}}^2A_{\mathrm{p}}^2}{2}e^{i(k_{\mathrm{p}}+k_{\mathrm{p}})x}-k_{\mathrm{i}}k_{\mathrm{p+s}}A_{\mathrm{i}}A_{\mathrm{p+s}}e^{i(k_{\mathrm{i}}+k_{\mathrm{p+s}})x}-k_{\mathrm{p+i}}k_{\mathrm{s}}A_{\mathrm{p+i}}A_{\mathrm{s}}e^{i(k_{\mathrm{p+i}}+k_{\mathrm{s}})x}\\
&+k_{\mathrm{p}}k_{\mathrm{3p}}A_{\mathrm{p}}^*A_{\mathrm{3p}}e^{i(k_{\mathrm{3p}}-k_{\mathrm{p}})x}+k_{\mathrm{2p+s}}k_{\mathrm{s}}A_{\mathrm{2p+s}}A_{\mathrm{s}}^*e^{i(k_{\mathrm{2p+s}}-k_{\mathrm{s}})x} \\
&+ k_{\mathrm{2p+i}}k_{\mathrm{i}}A_{\mathrm{2p+i}}A_{\mathrm{i}}^*e^{i(k_{\mathrm{2p+i}}-k_{\mathrm{i}})x} + k_{\mathrm{4p}}k_{\mathrm{2p+i}}A_{\mathrm{4p}}A_{\mathrm{2p}}^*e^{i(k_{\mathrm{4p}}-k_{\mathrm{2p}})x}\\
&+ k_{\mathrm{3p+s}}k_{\mathrm{p+s}}A_{\mathrm{3p+s}}A_{\mathrm{p+s}}^*e^{i(k_{\mathrm{3p+s}}-k_{\mathrm{p+s}})x}  + k_{\mathrm{3p+i}}k_{\mathrm{p+i}}A_{\mathrm{3p+i}}A_{\mathrm{p+i}}^*e^{i(k_{\mathrm{3p+i}}-k_{\mathrm{p+i}})x}\Big)e^{-ik_{\mathrm{2p}}x}
\label{eq:CME4-2p}
\end{aligned}\\
\dfrac{dA_{\mathrm{p+s}}}{dx} &=
\begin{aligned}[t]\dfrac{\beta}{2}\Big(&-k_{\mathrm{p}}k_{\mathrm{s}}A_{\mathrm{p}}A_{\mathrm{s}}e^{i(k_{\mathrm{p}}+k_{\mathrm{s}})x}+k_{\mathrm{i}}k_{\mathrm{2p}}A_{\mathrm{i}}^*A_{\mathrm{2p}}e^{i(k_{\mathrm{2p}}-k_{\mathrm{i}})x}\\
&+k_{\mathrm{3p}}k_{\mathrm{p+i}}A_{\mathrm{3p}}A_{\mathrm{p+i}}^*e^{i(k_{\mathrm{3p}}-k_{\mathrm{p+i}})x}+k_{\mathrm{2p+s}}k_{\mathrm{p}}A_{\mathrm{2p+s}}A_{\mathrm{p}}^*e^{i(k_{\mathrm{2p+s}}-k_{\mathrm{p}})x}\\
&+k_{\mathrm{4p}}k_{\mathrm{2p+i}}A_{\mathrm{4p}}A_{\mathrm{2p+i}}^*e^{i(k_{\mathrm{4p}}-k_{\mathrm{2p+i}})x} + k_{\mathrm{3p+s}}k_{\mathrm{2p}}A_{\mathrm{3p+s}}A_{\mathrm{2p}}^*e^{i(k_{\mathrm{3p+s}}-k_{\mathrm{2p}})x} \Big)e^{-ik_{\mathrm{p+s}}x}
\label{eq:CME4-p+s}
\end{aligned}\\
\dfrac{dA_{\mathrm{p+i}}}{dx} &=
\begin{aligned}[t]\dfrac{\beta}{2}\Big(&-k_{\mathrm{p}}k_{\mathrm{i}}A_{\mathrm{p}}A_{\mathrm{i}}e^{i(k_{\mathrm{p}}+k_{\mathrm{i}})x}+k_{\mathrm{s}}k_{\mathrm{2p}}A_{\mathrm{s}}^*A_{\mathrm{2p}}e^{i(k_{\mathrm{2p}}-k_{\mathrm{s}})x}\\
&  +k_{\mathrm{3p}}k_{\mathrm{p+s}}A_{\mathrm{3p}}A_{\mathrm{p+s}}^*e^{i(k_{\mathrm{3p}}-k_{\mathrm{p+s}})x}+k_{\mathrm{2p+i}}k_{\mathrm{p}}A_{\mathrm{2p+i}}A_{\mathrm{p}}^*e^{i(k_{\mathrm{2p+i}}-k_{\mathrm{p}})x}\\
& +k_{\mathrm{4p}}k_{\mathrm{2p+s}}A_{\mathrm{4p}}A_{\mathrm{2p+s}}^*e^{i(k_{\mathrm{4p}}-k_{\mathrm{2p+s}})x} + k_{\mathrm{3p+i}}k_{\mathrm{2p}}A_{\mathrm{3p+i}}A_{\mathrm{2p}}^*e^{i(k_{\mathrm{3p+i}}-k_{\mathrm{2p}})x}\Big)e^{-ik_{\mathrm{p+i}}x}
\label{eq:CME4-p+i}
\end{aligned}\\
\dfrac{dA_{\mathrm{3p}}}{dx} &=
\begin{aligned}[t]\dfrac{\beta}{2}\Big(&-k_{\mathrm{p}}k_{\mathrm{2p}}A_{\mathrm{p}}A_{\mathrm{2p}}e^{i(k_{\mathrm{p}}+k_{\mathrm{2p}})x}-k_{\mathrm{p+s}}k_{\mathrm{p+i}}A_{\mathrm{p+s}}A_{\mathrm{p+i}}e^{i(k_{\mathrm{p+s}}+k_{\mathrm{p+i}})x}   \\
&-k_{\mathrm{s}}k_{\mathrm{2p+i}}A_{\mathrm{s}}A_{\mathrm{2p+i}}e^{i(k_{\mathrm{s}}+k_{\mathrm{2p+i}})x}-k_{\mathrm{i}}k_{\mathrm{2p+s}}A_{\mathrm{i}}A_{\mathrm{2p+s}}e^{i(k_{\mathrm{i}}+k_{\mathrm{2p+s}})x}\\
&+k_{\mathrm{4p}}k_{\mathrm{p}}A_{\mathrm{4p}}A_{\mathrm{p}}^*e^{i(k_{\mathrm{4p}}-k_{\mathrm{p}})x} + k_{\mathrm{3p+s}}k_{\mathrm{s}}A_{\mathrm{3p+s}}A_{\mathrm{s}}^*e^{i(k_{\mathrm{3p+s}}-k_{\mathrm{s}})x}\\
&+k_{\mathrm{3p+i}}k_{\mathrm{i}}A_{\mathrm{3p+i}}A_{\mathrm{i}}^*e^{i(k_{\mathrm{3p+i}}-k_{\mathrm{i}})x}\Big)e^{-ik_{\mathrm{3p}}x}
\label{eq:CME4-3p}
\end{aligned}\\
\dfrac{dA_{\mathrm{2p+s}}}{dx} &=
\begin{aligned}[t]\dfrac{\beta}{2}\Big(&-k_{\mathrm{p}}k_{\mathrm{p+s}}A_{\mathrm{p}}A_{\mathrm{p+s}}e^{i(k_{\mathrm{p}}+k_{\mathrm{p+s}})x}-k_{\mathrm{s}}k_{\mathrm{2p}}A_{\mathrm{s}}A_{\mathrm{2p}}e^{i(k_{\mathrm{s}}+k_{\mathrm{2p}})x}\\
&+k_{\mathrm{3p}}k_{\mathrm{i}}A_{\mathrm{3p}}A_{\mathrm{i}}^*e^{i(k_{\mathrm{3p}}-k_{\mathrm{i}})x}+k_{\mathrm{4p}}k_{\mathrm{p+i}}A_{\mathrm{4p}}A_{\mathrm{p+i}}^*e^{i(k_{\mathrm{4p}}-k_{\mathrm{p+i}})x} \\
&+k_{\mathrm{3p+s}}k_{\mathrm{p}}A_{\mathrm{3p+s}}A_{\mathrm{p}}^*e^{i(k_{\mathrm{3p+s}}-k_{\mathrm{p}})x}\Big)e^{-ik_{\mathrm{2p+s}}x}
\label{eq:CME4-2p+s}
\end{aligned}\\
\dfrac{dA_{\mathrm{2p+i}}}{dx} &=
\begin{aligned}[t]\dfrac{\beta}{2}\Big(&-k_{\mathrm{p}}k_{\mathrm{p+i}}A_{\mathrm{p}}A_{\mathrm{p+i}}e^{i(k_{\mathrm{p}}+k_{\mathrm{p+i}})x}-k_{\mathrm{i}}k_{\mathrm{2p}}A_{\mathrm{i}}A_{\mathrm{2p}}e^{i(k_{\mathrm{i}}+k_{\mathrm{2p}})x}\\
&+k_{\mathrm{3p}}k_{\mathrm{s}}A_{\mathrm{3p}}A_{\mathrm{s}}^*e^{i(k_{\mathrm{3p}}-k_{\mathrm{s}})x}+k_{\mathrm{4p}}k_{\mathrm{p+s}}A_{\mathrm{4p}}A_{\mathrm{p+s}}^*e^{i(k_{\mathrm{4p}}-k_{\mathrm{p+s}})x} \\
&+ k_{\mathrm{3p+i}}k_{\mathrm{p}}A_{\mathrm{3p+i}}A_{\mathrm{p}}^*e^{i(k_{\mathrm{3p+i}}-k_{\mathrm{p}})x} \Big)e^{-ik_{\mathrm{2p+i}}x}
\label{eq:CME4-2p+i}
\end{aligned}\\
\dfrac{dA_{\mathrm{4p}}}{dx} &=
\begin{aligned}[t]\dfrac{\beta}{2}\Big(&-k_{\mathrm{p}}k_{\mathrm{3p}}A_{\mathrm{p}}A_{\mathrm{3p}}e^{i(k_{\mathrm{p}}+k_{\mathrm{3p}})x}-k_{\mathrm{s}}k_{\mathrm{3p+i}}A_{\mathrm{s}}A_{\mathrm{3p+i}}e^{i(k_{\mathrm{3p+i}}+k_{\mathrm{s}})x}\\
&-k_{\mathrm{i}}k_{\mathrm{3p+s}}A_{\mathrm{i}}A_{\mathrm{3p+s}}e^{i(k_{\mathrm{i}}+k_{\mathrm{3p+s}})x}-\dfrac{k_{\mathrm{2p}}^2A_{\mathrm{2p}}^2}{2}e^{i(k_{\mathrm{2p}}+k_{\mathrm{2p}})x}  \\
&-k_{\mathrm{p+s}}k_{\mathrm{2p+i}}A_{\mathrm{p+s}}A_{\mathrm{2p+i}}e^{i(k_{\mathrm{p+s}}+k_{\mathrm{2p+i}})x} - k_{\mathrm{p+i}}k_{\mathrm{2p+s}}A_{\mathrm{p+i}}A_{\mathrm{2p+s}}e^{i(k_{\mathrm{p+i}}+k_{\mathrm{2p+s}})x} \Big)e^{-ik_{\mathrm{4p}}x}
\label{eq:CME4-4p}
\end{aligned}
\\
\dfrac{dA_{\mathrm{3p+s}}}{dx} &=
\begin{aligned}[t]\dfrac{\beta}{2}\Big(&-k_{\mathrm{p}}k_{\mathrm{2p+s}}A_{\mathrm{p}}A_{\mathrm{2p+s}}e^{i(k_{\mathrm{p}}+k_{\mathrm{2p+s}})x}-k_{\mathrm{2p}}k_{\mathrm{p+s}}A_{\mathrm{2p}}A_{\mathrm{p+s}}e^{i(k_{\mathrm{2p}}+k_{\mathrm{p+s}})x}\\
&-k_{\mathrm{s}}k_{\mathrm{3p}}A_{\mathrm{s}}A_{\mathrm{3p}}e^{i(k_{\mathrm{s}}+k_{\mathrm{3p}})x}+k_{\mathrm{i}}k_{\mathrm{4p}}A_{\mathrm{i}}^*A_{\mathrm{4p}}e^{i(k_{\mathrm{4p}}-k_{\mathrm{i}})x} \Big)e^{-ik_{\mathrm{3p+s}}x}
\label{eq:CME4-3p+s}
\end{aligned}\\
\dfrac{dA_{\mathrm{3p+i}}}{dx} &=
\begin{aligned}[t]\dfrac{\beta}{2}\Big(&-k_{\mathrm{p}}k_{\mathrm{2p+i}}A_{\mathrm{p}}A_{\mathrm{2p+i}}e^{i(k_{\mathrm{p}}+k_{\mathrm{2p+i}})x}-k_{\mathrm{2p}}k_{\mathrm{p+i}}A_{\mathrm{2p}}A_{\mathrm{p+i}}e^{i(k_{\mathrm{2p}}+k_{\mathrm{p+i}})x}\\
&-k_{\mathrm{i}}k_{\mathrm{3p}}A_{\mathrm{i}}A_{\mathrm{3p}}e^{i(k_{\mathrm{i}}+k_{\mathrm{3p}})x}+k_{\mathrm{s}}k_{\mathrm{4p}}A_{\mathrm{s}}^*A_{\mathrm{4p}}e^{i(k_{\mathrm{4p}}-k_{\mathrm{s}})x} \Big)e^{-ik_{\mathrm{3p+i}}x}
\label{eq:CME4-3p+i}
\end{aligned}
\end{align}
\end{widetext}

\begin{widetext}
\section{Extension to CME-5}
\label{app:CME5}
The final coupled mode equation extension that we present in full is the extension from CME-4 to CME-5 by inclusion of the fifth harmonic of the pump $f_{\mathrm{5p}}$, and the sum-frequency terms $f_{\mathrm{4p+i}}$ and $f_{\mathrm{4p+s}}$. The full list of tones included in CME-5 is shown in \cref{tab:CME}. We show below in detail all of the terms included in the coupled mode equation forming CME-5.
\begin{align}
\dfrac{dA_{\mathrm{i}}}{dx} &=
\begin{aligned}[t]
\dfrac{\beta}{2}\Big(&k_{\mathrm{s}}k_{\mathrm{p}}A_{\mathrm{p}}A_{\mathrm{s}}^*e^{i(k_{\mathrm{p}}-k_{\mathrm{s}})x} + k_{\mathrm{p+i}}k_{\mathrm{p}}A_{\mathrm{p+i}}A_{\mathrm{p}}^*e^{i(k_{\mathrm{p+i}}-k_{\mathrm{p}})x} \\ 
&+ k_{\mathrm{2p}}k_{\mathrm{p+s}}A_{\mathrm{2p}}A_{\mathrm{p+s}}^*e^{i(k_{\mathrm{2p}}-k_{\mathrm{p+s}})x} + k_{\mathrm{2p+i}}k_{\mathrm{2p}}A_{\mathrm{2p+i}}A_{\mathrm{2p}}^*e^{i(k_{\mathrm{2p+i}}-k_{\mathrm{2p}})x} \\ 
&+ k_{\mathrm{3p}}k_{\mathrm{2p+s}}A_{\mathrm{3p}}A_{\mathrm{2p+s}}^*e^{i(k_{\mathrm{3p}}-k_{\mathrm{2p+s}})x} + k_{\mathrm{3p+i}}k_{\mathrm{3p}}A_{\mathrm{3p+i}}A_{\mathrm{3p}}^*e^{i(k_{\mathrm{3p+i}}-k_{\mathrm{3p}})x} \\ 
&+ k_{\mathrm{4p}}k_{\mathrm{3p+s}}A_{\mathrm{4p}}A_{\mathrm{3p+s}}^*e^{i(k_{\mathrm{4p}}-k_{\mathrm{3p+s}})x} + k_{\mathrm{4p+i}}k_{\mathrm{4p}}A_{\mathrm{4p+i}}A_{\mathrm{4p}}^*e^{i(k_{\mathrm{4p+i}}-k_{\mathrm{4p}})x} \\ 
&+ k_{\mathrm{5p}}k_{\mathrm{4p+s}}A_{\mathrm{5p}}A_{\mathrm{4p+s}}^*e^{i(k_{\mathrm{5p}}-k_{\mathrm{4p+s}})x}\Big)e^{-ik_{\mathrm{i}}x}
\label{eq:CME5-i}
\end{aligned}\\
\dfrac{dA_{\mathrm{s}}}{dx} &=
\begin{aligned}[t]
\dfrac{\beta}{2}\Big(&k_{\mathrm{i}}k_{\mathrm{p}}A_{\mathrm{p}}A_{\mathrm{i}}^*e^{i(k_{\mathrm{p}}-k_{\mathrm{i}})x} + k_{\mathrm{p+s}}k_{\mathrm{p}}A_{\mathrm{p+s}}A_{\mathrm{p}}^*e^{i(k_{\mathrm{p+s}}-k_{\mathrm{p}})x} \\
&+ k_{\mathrm{2p}}k_{\mathrm{p+i}}A_{\mathrm{2p}}A_{\mathrm{p+i}}^*e^{i(k_{\mathrm{2p}}-k_{\mathrm{p+i}})x}
+ k_{\mathrm{2p+s}}k_{\mathrm{2p}}A_{\mathrm{2p+s}}A_{\mathrm{2p}}^*e^{i(k_{\mathrm{2p+s}}-k_{\mathrm{2p}})x} \\
&+ k_{\mathrm{3p}}k_{\mathrm{2p+i}}A_{\mathrm{3p}}A_{\mathrm{2p+i}}^*e^{i(k_{\mathrm{3p}}-k_{\mathrm{2p+i}})x} + k_{\mathrm{3p+s}}k_{\mathrm{3p}}A_{\mathrm{3p+s}}A_{\mathrm{3p}}^*e^{i(k_{\mathrm{3p+s}}-k_{\mathrm{3p}})x} \\
&+ k_{\mathrm{4p}}k_{\mathrm{3p+i}}A_{\mathrm{4p}}A_{\mathrm{3p+i}}^*e^{i(k_{\mathrm{4p}}-k_{\mathrm{3p+i}})x} + k_{\mathrm{4p+s}}k_{\mathrm{4p}}A_{\mathrm{4p+s}}A_{\mathrm{4p}}^*e^{i(k_{\mathrm{4p+s}}-k_{\mathrm{4p}})x} \\
&+ k_{\mathrm{5p}}k_{\mathrm{4p+i}}A_{\mathrm{5p}}A_{\mathrm{4p+i}}^*e^{i(k_{\mathrm{5p}}-k_{\mathrm{4p+i}})x}\Big)e^{-ik_{\mathrm{s}}x}
\label{eq:CME5-s}
\end{aligned}\\
\dfrac{dA_{\mathrm{p}}}{dx} &=
\begin{aligned}[t]
\dfrac{\beta}{2}\Big(&-k_{\mathrm{i}}k_{\mathrm{s}}A_{\mathrm{i}}A_{\mathrm{s}}e^{i(k_{\mathrm{i}}+k_{\mathrm{s}})x} + k_{\mathrm{p+i}}k_{\mathrm{i}}A_{\mathrm{p+i}}A_{\mathrm{i}}^*e^{i(k_{\mathrm{p+i}}-k_{\mathrm{i}})x} \\
&+ k_{\mathrm{p+s}}k_{\mathrm{s}}A_{\mathrm{p+s}}A_{\mathrm{s}}^*e^{i(k_{\mathrm{p+s}}-k_{\mathrm{s}})x} + k_{\mathrm{2p}}k_{\mathrm{p}}A_{\mathrm{2p}}A_{\mathrm{p}}^*e^{i(k_{\mathrm{2p}}-k_{\mathrm{p}})x} \\
&+ k_{\mathrm{2p+i}}k_{\mathrm{p+i}}A_{\mathrm{2p+i}}A_{\mathrm{p+i}}^*e^{i(k_{\mathrm{2p+i}}-k_{\mathrm{p+i}})x} + k_{\mathrm{2p+s}}k_{\mathrm{p+s}}A_{\mathrm{2p+s}}A_{\mathrm{p+s}}^*e^{i(k_{\mathrm{2p+s}}-k_{\mathrm{p+s}})x} \\
&+ k_{\mathrm{3p}}k_{\mathrm{2p}}A_{\mathrm{3p}}A_{\mathrm{2p}}^*e^{i(k_{\mathrm{3p}}-k_{\mathrm{2p}})x} + k_{\mathrm{3p+i}}k_{\mathrm{2p+i}}A_{\mathrm{3p+i}}A_{\mathrm{2p+i}}^*e^{i(k_{\mathrm{3p+i}}-k_{\mathrm{2p+i}})x} \\
&+ k_{\mathrm{3p+s}}k_{\mathrm{2p+s}}A_{\mathrm{3p+s}}A_{\mathrm{2p+s}}^*e^{i(k_{\mathrm{3p+s}}-k_{\mathrm{2p+s}})x} + k_{\mathrm{4p}}k_{\mathrm{3p}}A_{\mathrm{4p}}A_{\mathrm{3p}}^*e^{i(k_{\mathrm{4p}}-k_{\mathrm{3p}})x} \\
&+ k_{\mathrm{4p+i}}k_{\mathrm{3p+i}}A_{\mathrm{4p+i}}A_{\mathrm{3p+i}}^*e^{i(k_{\mathrm{4p+i}}-k_{\mathrm{3p+i}})x} + k_{\mathrm{4p+s}}k_{\mathrm{3p+s}}A_{\mathrm{4p+s}}A_{\mathrm{3p+s}}^*e^{i(k_{\mathrm{4p+s}}-k_{\mathrm{3p+s}})x} \\
&+ k_{\mathrm{5p}}k_{\mathrm{4p}}A_{\mathrm{5p}}A_{\mathrm{4p}}^*e^{i(k_{\mathrm{5p}}-k_{\mathrm{4p}})x}\Big){2}e^{-ik_{\mathrm{p}}x}
\label{eq:CME5-p}
\end{aligned}\\
\dfrac{dA_{\mathrm{p+i}}}{dx} &=
\begin{aligned}[t]
\dfrac{\beta}{2}\Big(&-k_{\mathrm{i}}k_{\mathrm{p}}A_{\mathrm{i}}A_{\mathrm{p}}e^{i(k_{\mathrm{i}}+k_{\mathrm{p}})x} + k_{\mathrm{2p}}k_{\mathrm{s}}A_{\mathrm{2p}}A_{\mathrm{s}}^*e^{i(k_{\mathrm{2p}}-k_{\mathrm{s}})x} \\
&+ k_{\mathrm{2p+i}}k_{\mathrm{p}}A_{\mathrm{2p+i}}A_{\mathrm{p}}^*e^{i(k_{\mathrm{2p+i}}-k_{\mathrm{p}})x} + k_{\mathrm{3p}}k_{\mathrm{p+s}}A_{\mathrm{3p}}A_{\mathrm{p+s}}^*e^{i(k_{\mathrm{3p}}-k_{\mathrm{p+s}})x} \\
&+ k_{\mathrm{3p+i}}k_{\mathrm{2p}}A_{\mathrm{3p+i}}A_{\mathrm{2p}}^*e^{i(k_{\mathrm{3p+i}}-k_{\mathrm{2p}})x} + k_{\mathrm{4p}}k_{\mathrm{2p+s}}A_{\mathrm{4p}}A_{\mathrm{2p+s}}^*e^{i(k_{\mathrm{4p}}-k_{\mathrm{2p+s}})x} \\
&+ k_{\mathrm{4p+i}}k_{\mathrm{3p}}A_{\mathrm{4p+i}}A_{\mathrm{3p}}^*e^{i(k_{\mathrm{4p+i}}-k_{\mathrm{3p}})x} + k_{\mathrm{5p}}k_{\mathrm{3p+s}}A_{\mathrm{5p}}A_{\mathrm{3p+s}}^*e^{i(k_{\mathrm{5p}}-k_{\mathrm{3p+s}})x}\Big)e^{-ik_{\mathrm{p+i}}x}
\label{eq:CME5-p+i}
\end{aligned}\\
\dfrac{dA_{\mathrm{p+s}}}{dx} &=
\begin{aligned}[t]
\dfrac{\beta}{2}\Big(&-k_{\mathrm{s}}k_{\mathrm{p}}A_{\mathrm{s}}A_{\mathrm{p}}e^{i(k_{\mathrm{s}}+k_{\mathrm{p}})x} + k_{\mathrm{2p}}k_{\mathrm{i}}A_{\mathrm{2p}}A_{\mathrm{i}}^*e^{i(k_{\mathrm{2p}}-k_{\mathrm{i}})x} \\
&+ k_{\mathrm{2p+s}}k_{\mathrm{p}}A_{\mathrm{2p+s}}A_{\mathrm{p}}^*e^{i(k_{\mathrm{2p+s}}-k_{\mathrm{p}})x} + k_{\mathrm{3p}}k_{\mathrm{p+i}}A_{\mathrm{3p}}A_{\mathrm{p+i}}^*e^{i(k_{\mathrm{3p}}-k_{\mathrm{p+i}})x} \\
&+ k_{\mathrm{3p+s}}k_{\mathrm{2p}}A_{\mathrm{3p+s}}A_{\mathrm{2p}}^*e^{i(k_{\mathrm{3p+s}}-k_{\mathrm{2p}})x} + k_{\mathrm{4p}}k_{\mathrm{2p+i}}A_{\mathrm{4p}}A_{\mathrm{2p+i}}^*e^{i(k_{\mathrm{4p}}-k_{\mathrm{2p+i}})x} \\
&+ k_{\mathrm{4p+s}}k_{\mathrm{3p}}A_{\mathrm{4p+s}}A_{\mathrm{3p}}^*e^{i(k_{\mathrm{4p+s}}-k_{\mathrm{3p}})x} + k_{\mathrm{5p}}k_{\mathrm{3p+i}}A_{\mathrm{5p}}A_{\mathrm{3p+i}}^*e^{i(k_{\mathrm{5p}}-k_{\mathrm{3p+i}})x}\Big)e^{-ik_{\mathrm{p+s}}x}
\label{eq:CME5-p+s}
\end{aligned}\\
\dfrac{dA_{\mathrm{2p}}}{dx} &=
\begin{aligned}[t]
\dfrac{\beta}{2}\Big(&-k_{\mathrm{i}}k_{\mathrm{p+s}}A_{\mathrm{i}}A_{\mathrm{p+s}}e^{i(k_{\mathrm{i}}+k_{\mathrm{p+s}})x} - k_{\mathrm{s}}k_{\mathrm{p+i}}A_{\mathrm{s}}A_{\mathrm{p+i}}e^{i(k_{\mathrm{s}}+k_{\mathrm{p+i}})x} \\
&- \dfrac{k_{\mathrm{p}}^2A_{\mathrm{p}}^2}{2}e^{i(k_{\mathrm{p}}+k_{\mathrm{p}})x} + k_{\mathrm{2p+i}}k_{\mathrm{i}}A_{\mathrm{2p+i}}A_{\mathrm{i}}^*e^{i(k_{\mathrm{2p+i}}-k_{\mathrm{i}})x} \\
&+ k_{\mathrm{2p+s}}k_{\mathrm{s}}A_{\mathrm{2p+s}}A_{\mathrm{s}}^*e^{i(k_{\mathrm{2p+s}}-k_{\mathrm{s}})x} + k_{\mathrm{3p}}k_{\mathrm{p}}A_{\mathrm{3p}}A_{\mathrm{p}}^*e^{i(k_{\mathrm{3p}}-k_{\mathrm{p}})x} \\
&+ k_{\mathrm{3p+i}}k_{\mathrm{p+i}}A_{\mathrm{3p+i}}A_{\mathrm{p+i}}^*e^{i(k_{\mathrm{3p+i}}-k_{\mathrm{p+i}})x} + k_{\mathrm{3p+s}}k_{\mathrm{p+s}}A_{\mathrm{3p+s}}A_{\mathrm{p+s}}^*e^{i(k_{\mathrm{3p+s}}-k_{\mathrm{p+s}})x} \\
&+ k_{\mathrm{4p}}k_{\mathrm{2p}}A_{\mathrm{4p}}A_{\mathrm{2p}}^*e^{i(k_{\mathrm{4p}}-k_{\mathrm{2p}})x} + k_{\mathrm{4p+i}}k_{\mathrm{2p+i}}A_{\mathrm{4p+i}}A_{\mathrm{2p+i}}^*e^{i(k_{\mathrm{4p+i}}-k_{\mathrm{2p+i}})x} \\
&+ k_{\mathrm{4p+s}}k_{\mathrm{2p+s}}A_{\mathrm{4p+s}}A_{\mathrm{2p+s}}^*e^{i(k_{\mathrm{4p+s}}-k_{\mathrm{2p+s}})x} + k_{\mathrm{5p}}k_{\mathrm{3p}}A_{\mathrm{5p}}A_{\mathrm{3p}}^*e^{i(k_{\mathrm{5p}}-k_{\mathrm{3p}})x}\Big)e^{-ik_{\mathrm{2p}}x}
\label{eq:CME5-2p}
\end{aligned}\\
\dfrac{dA_{\mathrm{2p+i}}}{dx} &=
\begin{aligned}[t]
\dfrac{\beta}{2}\Big(&-k_{\mathrm{i}}k_{\mathrm{2p}}A_{\mathrm{i}}A_{\mathrm{2p}}e^{i(k_{\mathrm{i}}+k_{\mathrm{2p}})x} - k_{\mathrm{p}}k_{\mathrm{p+i}}A_{\mathrm{p}}A_{\mathrm{p+i}}e^{i(k_{\mathrm{p}}+k_{\mathrm{p+i}})x} \\
&+ k_{\mathrm{3p}}k_{\mathrm{s}}A_{\mathrm{3p}}A_{\mathrm{s}}^*e^{i(k_{\mathrm{3p}}-k_{\mathrm{s}})x} + k_{\mathrm{3p+i}}k_{\mathrm{p}}A_{\mathrm{3p+i}}A_{\mathrm{p}}^*e^{i(k_{\mathrm{3p+i}}-k_{\mathrm{p}})x} \\
&+ k_{\mathrm{4p}}k_{\mathrm{p+s}}A_{\mathrm{4p}}A_{\mathrm{p+s}}^*e^{i(k_{\mathrm{4p}}-k_{\mathrm{p+s}})x} + k_{\mathrm{4p+i}}k_{\mathrm{2p}}A_{\mathrm{4p+i}}A_{\mathrm{2p}}^*e^{i(k_{\mathrm{4p+i}}-k_{\mathrm{2p}})x} \\
&+ k_{\mathrm{5p}}k_{\mathrm{2p+s}}A_{\mathrm{5p}}A_{\mathrm{2p+s}}^*e^{i(k_{\mathrm{5p}}-k_{\mathrm{2p+s}})x}\Big)e^{-ik_{\mathrm{2p+i}}x}
\label{eq:CME5-2p+i}
\end{aligned}\\
\dfrac{dA_{\mathrm{2p+s}}}{dx} &=
\begin{aligned}[t]
\dfrac{\beta}{2}\Big(&-k_{\mathrm{s}}k_{\mathrm{2p}}A_{\mathrm{s}}A_{\mathrm{2p}}e^{i(k_{\mathrm{s}}+k_{\mathrm{2p}})x} - k_{\mathrm{p}}k_{\mathrm{p+s}}A_{\mathrm{p}}A_{\mathrm{p+s}}e^{i(k_{\mathrm{p}}+k_{\mathrm{p+s}})x} \\
&+ k_{\mathrm{3p}}k_{\mathrm{i}}A_{\mathrm{3p}}A_{\mathrm{i}}^*e^{i(k_{\mathrm{3p}}-k_{\mathrm{i}})x} + k_{\mathrm{3p+s}}k_{\mathrm{p}}A_{\mathrm{3p+s}}A_{\mathrm{p}}^*e^{i(k_{\mathrm{3p+s}}-k_{\mathrm{p}})x} \\
&+ k_{\mathrm{4p}}k_{\mathrm{p+i}}A_{\mathrm{4p}}A_{\mathrm{p+i}}^*e^{i(k_{\mathrm{4p}}-k_{\mathrm{p+i}})x} + k_{\mathrm{4p+s}}k_{\mathrm{2p}}A_{\mathrm{4p+s}}A_{\mathrm{2p}}^*e^{i(k_{\mathrm{4p+s}}-k_{\mathrm{2p}})x} \\
&+ k_{\mathrm{5p}}k_{\mathrm{2p+i}}A_{\mathrm{5p}}A_{\mathrm{2p+i}}^*e^{i(k_{\mathrm{5p}}-k_{\mathrm{2p+i}})x}\Big)e^{-ik_{\mathrm{2p+s}}x}
\label{eq:CME5-2p+s}
\end{aligned}\\
\dfrac{dA_{\mathrm{3p}}}{dx} &=
\begin{aligned}[t]
\dfrac{\beta}{2}\Big(&-k_{\mathrm{i}}k_{\mathrm{2p+s}}A_{\mathrm{i}}A_{\mathrm{2p+s}}e^{i(k_{\mathrm{i}}+k_{\mathrm{2p+s}})x} - k_{\mathrm{s}}k_{\mathrm{2p+i}}A_{\mathrm{s}}A_{\mathrm{2p+i}}e^{i(k_{\mathrm{s}}+k_{\mathrm{2p+i}})x} \\
&- k_{\mathrm{p}}k_{\mathrm{2p}}A_{\mathrm{p}}A_{\mathrm{2p}}e^{i(k_{\mathrm{p}}+k_{\mathrm{2p}})x} - k_{\mathrm{p+i}}k_{\mathrm{p+s}}A_{\mathrm{p+i}}A_{\mathrm{p+s}}e^{i(k_{\mathrm{p+i}}+k_{\mathrm{p+s}})x} \\
&+ k_{\mathrm{3p+i}}k_{\mathrm{i}}A_{\mathrm{3p+i}}A_{\mathrm{i}}^*e^{i(k_{\mathrm{3p+i}}-k_{\mathrm{i}})x} + k_{\mathrm{3p+s}}k_{\mathrm{s}}A_{\mathrm{3p+s}}A_{\mathrm{s}}^*e^{i(k_{\mathrm{3p+s}}-k_{\mathrm{s}})x} \\
&+ k_{\mathrm{4p}}k_{\mathrm{p}}A_{\mathrm{4p}}A_{\mathrm{p}}^*e^{i(k_{\mathrm{4p}}-k_{\mathrm{p}})x} + k_{\mathrm{4p+i}}k_{\mathrm{p+i}}A_{\mathrm{4p+i}}A_{\mathrm{p+i}}^*e^{i(k_{\mathrm{4p+i}}-k_{\mathrm{p+i}})x} \\
&+ k_{\mathrm{4p+s}}k_{\mathrm{p+s}}A_{\mathrm{4p+s}}A_{\mathrm{p+s}}^*e^{i(k_{\mathrm{4p+s}}-k_{\mathrm{p+s}})x} + k_{\mathrm{5p}}k_{\mathrm{2p}}A_{\mathrm{5p}}A_{\mathrm{2p}}^*e^{i(k_{\mathrm{5p}}-k_{\mathrm{2p}})x}\Big)e^{-ik_{\mathrm{3p}}x}
\label{eq:CME5-3p}
\end{aligned}\\
\dfrac{dA_{\mathrm{3p+i}}}{dx} &=
\begin{aligned}[t]
\dfrac{\beta}{2}\Big(&-k_{\mathrm{i}}k_{\mathrm{3p}}A_{\mathrm{i}}A_{\mathrm{3p}}e^{i(k_{\mathrm{i}}+k_{\mathrm{3p}})x} - k_{\mathrm{p}}k_{\mathrm{2p+i}}A_{\mathrm{p}}A_{\mathrm{2p+i}}e^{i(k_{\mathrm{p}}+k_{\mathrm{2p+i}})x} \\
&- k_{\mathrm{p+i}}k_{\mathrm{2p}}A_{\mathrm{p+i}}A_{\mathrm{2p}}e^{i(k_{\mathrm{p+i}}+k_{\mathrm{2p}})x} + k_{\mathrm{4p}}k_{\mathrm{s}}A_{\mathrm{4p}}A_{\mathrm{s}}^*e^{i(k_{\mathrm{4p}}-k_{\mathrm{s}})x} \\
&+ k_{\mathrm{4p+i}}k_{\mathrm{p}}A_{\mathrm{4p+i}}A_{\mathrm{p}}^*e^{i(k_{\mathrm{4p+i}}-k_{\mathrm{p}})x} + k_{\mathrm{5p}}k_{\mathrm{p+s}}A_{\mathrm{5p}}A_{\mathrm{p+s}}^*e^{i(k_{\mathrm{5p}}-k_{\mathrm{p+s}})x})e^{-ik_{\mathrm{3p+i}}x}
\label{eq:CME5-3p+i}
\end{aligned}\\
\dfrac{dA_{\mathrm{3p+s}}}{dx} &=
\begin{aligned}[t]
\dfrac{\beta}{2}\Big(&-k_{\mathrm{s}}k_{\mathrm{3p}}A_{\mathrm{s}}A_{\mathrm{3p}}e^{i(k_{\mathrm{s}}+k_{\mathrm{3p}})x} - k_{\mathrm{p}}k_{\mathrm{2p+s}}A_{\mathrm{p}}A_{\mathrm{2p+s}}e^{i(k_{\mathrm{p}}+k_{\mathrm{2p+s}})x} \\
&- k_{\mathrm{p+s}}k_{\mathrm{2p}}A_{\mathrm{p+s}}A_{\mathrm{2p}}e^{i(k_{\mathrm{p+s}}+k_{\mathrm{2p}})x} + k_{\mathrm{4p}}k_{\mathrm{i}}A_{\mathrm{4p}}A_{\mathrm{i}}^*e^{i(k_{\mathrm{4p}}-k_{\mathrm{i}})x} \\
&+ k_{\mathrm{4p+s}}k_{\mathrm{p}}A_{\mathrm{4p+s}}A_{\mathrm{p}}^*e^{i(k_{\mathrm{4p+s}}-k_{\mathrm{p}})x} + k_{\mathrm{5p}}k_{\mathrm{p+i}}A_{\mathrm{5p}}A_{\mathrm{p+i}}^*e^{i(k_{\mathrm{5p}}-k_{\mathrm{p+i}})x}\Big)e^{-ik_{\mathrm{3p+s}}x}
\label{eq:CME5-3p+s}
\end{aligned}\\
\dfrac{dA_{\mathrm{4p}}}{dx} &=
\begin{aligned}[t]
\dfrac{\beta}{2}\Big(&-k_{\mathrm{i}}k_{\mathrm{3p+s}}A_{\mathrm{i}}A_{\mathrm{3p+s}}e^{i(k_{\mathrm{i}}+k_{\mathrm{3p+s}})x} - k_{\mathrm{s}}k_{\mathrm{3p+i}}A_{\mathrm{s}}A_{\mathrm{3p+i}}e^{i(k_{\mathrm{s}}+k_{\mathrm{3p+i}})x} \\
&- k_{\mathrm{p}}k_{\mathrm{3p}}A_{\mathrm{p}}A_{\mathrm{3p}}e^{i(k_{\mathrm{p}}+k_{\mathrm{3p}})x} - k_{\mathrm{p+i}}k_{\mathrm{2p+s}}A_{\mathrm{p+i}}A_{\mathrm{2p+s}}e^{i(k_{\mathrm{p+i}}+k_{\mathrm{2p+s}})x} \\
&- k_{\mathrm{p+s}}k_{\mathrm{2p+i}}A_{\mathrm{p+s}}A_{\mathrm{2p+i}}e^{i(k_{\mathrm{p+s}}+k_{\mathrm{2p+i}})x} - \dfrac{k_{\mathrm{2p}}^2A_{\mathrm{2p}}^2}{2}e^{i(k_{\mathrm{2p}}+k_{\mathrm{2p}})x} \\
&+ k_{\mathrm{4p+i}}k_{\mathrm{i}}A_{\mathrm{4p+i}}A_{\mathrm{i}}^*e^{i(k_{\mathrm{4p+i}}-k_{\mathrm{i}})x} + k_{\mathrm{4p+s}}k_{\mathrm{s}}A_{\mathrm{4p+s}}A_{\mathrm{s}}^*e^{i(k_{\mathrm{4p+s}}-k_{\mathrm{s}})x} \\
&+ k_{\mathrm{5p}}k_{\mathrm{p}}A_{\mathrm{5p}}A_{\mathrm{p}}^*e^{i(k_{\mathrm{5p}}-k_{\mathrm{p}})x}\Big)e^{-ik_{\mathrm{4p}}x}
\label{eq:CME5-4p}
\end{aligned}\\
\dfrac{dA_{\mathrm{4p+i}}}{dx} &=
\begin{aligned}[t]
\dfrac{\beta}{2}\Big(&-k_{\mathrm{i}}k_{\mathrm{4p}}A_{\mathrm{i}}A_{\mathrm{4p}}e^{i(k_{\mathrm{i}}+k_{\mathrm{4p}})x} - k_{\mathrm{p}}k_{\mathrm{3p+i}}A_{\mathrm{p}}A_{\mathrm{3p+i}}e^{i(k_{\mathrm{p}}+k_{\mathrm{3p+i}})x} \\
&- k_{\mathrm{p+i}}k_{\mathrm{3p}}A_{\mathrm{p+i}}A_{\mathrm{3p}}e^{i(k_{\mathrm{p+i}}+k_{\mathrm{3p}})x} - k_{\mathrm{2p}}k_{\mathrm{2p+i}}A_{\mathrm{2p}}A_{\mathrm{2p+i}}e^{i(k_{\mathrm{2p}}+k_{\mathrm{2p+i}})x} \\
&+ k_{\mathrm{5p}}k_{\mathrm{s}}A_{\mathrm{5p}}A_{\mathrm{s}}^*e^{i(k_{\mathrm{5p}}-k_{\mathrm{s}})x}\Big)e^{i(-k_{\mathrm{4p+i}})x}
\label{eq:CME5-4p+i}
\end{aligned}\\
\dfrac{dA_{\mathrm{4p+s}}}{dx} &=
\begin{aligned}[t]
\dfrac{\beta}{2}\Big(&-k_{\mathrm{s}}k_{\mathrm{4p}}A_{\mathrm{s}}A_{\mathrm{4p}}e^{i(k_{\mathrm{s}}+k_{\mathrm{4p}})x} - k_{\mathrm{p}}k_{\mathrm{3p+s}}A_{\mathrm{p}}A_{\mathrm{3p+s}}e^{i(k_{\mathrm{p}}+k_{\mathrm{3p+s}})x} \\
&- k_{\mathrm{p+s}}k_{\mathrm{3p}}A_{\mathrm{p+s}}A_{\mathrm{3p}}e^{i(k_{\mathrm{p+s}}+k_{\mathrm{3p}})x} - k_{\mathrm{2p}}k_{\mathrm{2p+s}}A_{\mathrm{2p}}A_{\mathrm{2p+s}}e^{i(k_{\mathrm{2p}}+k_{\mathrm{2p+s}})x} \\
&+ k_{\mathrm{5p}}k_{\mathrm{i}}A_{\mathrm{5p}}A_{\mathrm{i}}^*e^{i(k_{\mathrm{5p}}-k_{\mathrm{i}})x}\Big)e^{-ik_{\mathrm{4p+s}}x}\label{eq:CME5-4p+s}
\end{aligned}\\
\dfrac{dA_{\mathrm{5p}}}{dx} &=
\begin{aligned}[t]
\dfrac{\beta}{2}\Big(&-k_{\mathrm{i}}k_{\mathrm{4p+s}}A_{\mathrm{i}}A_{\mathrm{4p+s}}e^{i(k_{\mathrm{i}}+k_{\mathrm{4p+s}})x} - k_{\mathrm{s}}k_{\mathrm{4p+i}}A_{\mathrm{s}}A_{\mathrm{4p+i}}e^{i(k_{\mathrm{s}}+k_{\mathrm{4p+i}})x} \\
&- k_{\mathrm{p}}k_{\mathrm{4p}}A_{\mathrm{p}}A_{\mathrm{4p}}e^{i(k_{\mathrm{p}}+k_{\mathrm{4p}})x} - k_{\mathrm{p+i}}k_{\mathrm{3p+s}}A_{\mathrm{p+i}}A_{\mathrm{3p+s}}e^{i(k_{\mathrm{p+i}}+k_{\mathrm{3p+s}})x} \\
&- k_{\mathrm{p+s}}k_{\mathrm{3p+i}}A_{\mathrm{p+s}}A_{\mathrm{3p+i}}e^{i(k_{\mathrm{p+s}}+k_{\mathrm{3p+i}})x} - k_{\mathrm{2p}}k_{\mathrm{3p}}A_{\mathrm{2p}}A_{\mathrm{3p}}e^{i(k_{\mathrm{2p}}+k_{\mathrm{3p}})x} \\
&- k_{\mathrm{2p+i}}k_{\mathrm{2p+s}}A_{\mathrm{2p+i}}A_{\mathrm{2p+s}}e^{i(k_{\mathrm{2p+i}}+k_{\mathrm{2p+s}})x}\Big)e^{-ik_{\mathrm{5p}}x}
\label{eq:CME5-5p}
\end{aligned}
\end{align}
\end{widetext}

\bibliography{Parametric_bibliography}

\end{document}